\documentclass[journal]{IEEEtran}
\usepackage{mathrsfs}
\usepackage{amsfonts}
\usepackage{balance}
\usepackage{ifpdf}
\usepackage{graphicx}
\usepackage{cite}
\usepackage[cmex10]{amsmath}
\usepackage{algorithmic}
\usepackage{array}
\usepackage{threeparttable}
\usepackage{multirow}
\usepackage{subfigure}
\usepackage{mdwmath}
\usepackage{mdwtab}
\usepackage{booktabs}
\usepackage{bm}
\usepackage{hhline}
\usepackage[all]{xy}
\usepackage{color}
\usepackage{amssymb}
\usepackage[justification=centering]{caption}
\allowdisplaybreaks[4]
\usepackage{stfloats}
\begin{document}

\title{Asymptotically Optimal Sequence Sets With Low/Zero Ambiguity Zone Properties}

\author{Liying Tian,~\IEEEmembership{Member, IEEE}, Xiaoshi Song,~\IEEEmembership{Member, IEEE}, Zilong Liu,~\IEEEmembership{Senior Member, IEEE}, and Yubo Li,~\IEEEmembership{Member, IEEE} 

\thanks{Liying Tian and Xiaoshi Song are with the School of Computer Science and Engineering, Northeastern University, Shenyang 110819, China, and also with the Key Laboratory of Intelligent Computing in Medical Image, Ministry of Education, Northeastern University, Shenyang 110189, China (e-mail: tianliying@mail.neu.edu.cn; songxiaoshi@cse.neu.edu.cn) }

\thanks{Zilong Liu is with the School of Computer Science and Electronics Engineering, University of Essex, Colchester CO4 3SQ, U.K. (e-mail: zilong.liu@essex.ac.uk).}

\thanks{Yubo Li is with the School of Information Science and Engineering, Yanshan University, Qinhuangdao 066004, China, and also with the Hebei Key Laboratory of Information Transmission and Signal Processing, Qinhuangdao 066004, China (e-mail: liyubo6316@ysu.edu.cn). }}

\maketitle
\begin{abstract}
Sequences with low/zero ambiguity zone (LAZ/ZAZ) properties are useful in modern communication and radar systems operating over mobile environments.
This paper first presents a new family of ZAZ sequence sets motivated by the ``modulating'' zero correlation zone (ZCZ) sequences which were first proposed by Popovic and Mauritz.
We then introduce a second family of ZAZ sequence sets with comb-like spectrum, whereby the local Doppler resilience is guaranteed by their inherent spectral nulls in the frequency domain.
Finally, LAZ sequence sets are obtained by exploiting their connection with a novel class of mapping functions.
These proposed unimodular ZAZ and LAZ sequence sets are cyclically distinct and asymptotically optimal with respect to the existing theoretical bounds on ambiguity functions.
\end{abstract}


\IEEEpeerreviewmaketitle

\section{Introduction}

\IEEEPARstart{S}{equences} with good correlation properties are desirable in wireless communication and radar systems for a number of applications, such as synchronization, channel estimation, multiuser communication, interference mitigation, sensing, ranging, and positioning [1].
According to the Welch bound, however, it is impossible to obtain a sequence set having both ideal auto- and cross-correlation properties [2].
To circumvent this problem, extensive studies have been conducted on low-correlation sequences and low/zero correlation zone (LCZ/ZCZ) sequences, where the latter are characterized by low/zero correlation properties within a time-shift zone around the origin [3], [4].

Modern sequence design is more stringent as one is expected to deal with the notorious Doppler effect in various mobile  channels [5]-[7].
For example, in Vehicle-to-Everything (V2X) networks, satellite communications, as well as radar sensing systems, the received signals are often corrupted by both time delays and phase rotations introduced by the propagation delay and mobility-incurred Doppler, respectively.
To characterize the delay-Doppler response at the receiver side, ambiguity function is widely used [8].
For reliable estimation of the delay and Doppler values, it is required to minimize the auto-ambiguity sidelobes and cross-ambiguity magnitudes of a sequence set over the entire delay-Doppler domain. Unfortunately, such a design task is challenging.
An explicit algorithm was developed in [9] to generate a sequence set with low ambiguity property (called a finite oscillator system) from the Weil representation.
Then, Wang and Gong constructed in [10]-[12] several classes of complex-valued sequence sets with low ambiguity amplitudes using additive and multiplicative characters over finite fields.
Ding \emph{et al.} [13] introduced a set of ambiguity function bounds for unimodular sequence sets as well as four classes of unimodular sequence sets with good ambiguity properties.
Recently, a generic cubic-phase sequence set was introduced in [14], whereby each sequence possesses optimal low auto-ambiguity sidelobes and distinct sequences have low cross-ambiguity magnitudes.
To date, however, the construction of a sequence set with optimal auto- and cross-ambiguity properties is largely open.

\begin{table*}[t]
\caption{Comparison of periodic unimodular LAZ/ZAZ sequence sets}
\centering
\begin{threeparttable}
\setlength{\tabcolsep}{0.1em}
{\begin{tabular}{|c|c|c|c|c|c|c|c|c|}
\hline
\multicolumn{2}{|c|}{Method} &Length &Set size &$\theta _{\textrm{max}}$ &$Z_x$ &$Z_y$ &Constraint &Optimality\\
\hline
\multirow{5}{*}{{[13]}}&\emph{Theorem 9} &$L=q-1$    &1  &$\sqrt{L}$   &$L$ &$L$ &$q=p^l$, $p$ is a prime &optimal\\
\cline{2-9}
&\emph{Theorem 10} &$L=\frac{q-1}{N}$ &$N$ &$\sqrt{q}$ &$L$ &$L$ &$q=p^l$, $p$ is a prime, $N|(q-1)$, $N\geq 2$ &\\
\cline{2-9}
&\multirow{4}{*}{\emph{Theorem 11}} &{$L=\frac{q-1}{N}$} &{$L$} &{$2\sqrt{q}$} &{$L$} &{$L$} &{$q=p^l$, $p$ is a prime, $N|(q-1)$, $N\geq 2$} &\\
\cline{3-9}
& &\multirow{3}{*}{$L=\frac{q-1}{N}$} &\multirow{3}{*}{$\prod_{j=1}^s(\frac{L}{(L.\lambda_i)}+1)$} &\multirow{3}{*}{$\frac{\lambda_s\sqrt{q}}{L}$} &\multirow{3}{*}{$L$} &\multirow{3}{*}{$L$} &$q=p^l$, $p$ is a prime, $N|(q-1)$, $N\geq 2$,&\\
&&&&&&& $\{\lambda_i\}_{i=0}^\infty$  are positive integers coprime& \\
&&&&&&&to $q$ in increasing order& \\\hline
\multirow{4}{*}{[14]}
&{\emph{Theorem 5}} &$p$ &$p^2$ &$2\sqrt{p}$ &$p$ &$p$ &$p$ is an odd prime &\\
\cline{2-9}
&{\emph{Theorem 6}} &$p$ &$1$ &$\sqrt{p}$ &$p$ &$p$ &$p$ is an odd prime &optimal\\
\cline{2-9}
&{\emph{Theorem 8}} &$L$ &$1$ &$0$ &$\frac{L}{r}$ &$r$ &$\text{gcd}(a,L)=1$  if $L$ is odd, $r=\text{gcd}(2a,L)$, $r>1$ &optimal\\
\cline{2-9}
&\multirow{2}{*}{\emph{Construction 2}} &\multirow{2}{*}{$L$} &\multirow{2}{*}{$N$} &\multirow{2}{*}{0} &\multirow{2}{*}{$\frac{\left\lfloor{L}/{N} \right\rfloor }{r}$} &\multirow{2}{*}{$r$}&\multirow{2}{*}{$\text{gcd}(a,L)=1$ if $L$ is odd, $r=\text{gcd}(2a,L)$, $r>1$}&optimal\\
&&&&&&&&  if $N|L$\\
 \hline
\multirow{5}{*}{[26]} &\multirow{5}{*}{\emph{Theorem 4}} &\multirow{5}{*}{$2(2^m-1)$} &\multirow{5}{*}{$2M$}&\multirow{5}{*}{$2\sqrt{2^m}$} &\multirow{5}{*}{$L$}&\multirow{5}{*}{$2(2^m-1)$}
&$E = \{ {\bm{e}_i}=(e_{i,0},e_{i,1}):0 \le i < M \}$ is a shift &\\
&&&&&&& sequence  set, &\\
&&&&&&&$L=\min \big\{ \mathop {\min}\limits_{{\bm{e}} \ne {\bm{h}} \in E} \{2(e_0-h_0),2(e_1- h_1)\}, $ &\\
&&&&&&& $\mathop {\min}\limits_{\bm{e} \ne \bm{h} \in E} \{ 2(e_0-h_1)+ 1,2(e_1-h_0)-1 \} \big\}$ &\\
\hline
\multirow{6}{*}{This paper}&\multirow{2}{*}{\emph{Corollary 1}} &\multirow{2}{*}{$M{{N}^{2}}$} &\multirow{2}{*}{$MN$} &\multirow{2}{*}{0} &\multirow{2}{*}{$\left\lfloor \frac{N}{K} \right\rfloor$} &\multirow{2}{*}{$K$}  &\multirow{2}{*}{$K<N$, ${\rm{gcd}}(K,N)=1$} &{asymptotically}\\
&&&&&&&&{optimal}\\
\cline{2-9}
\cline{2-9}
&\multirow{2}{*}{\emph{Theorem 3}} &\multirow{2}{*}{$N(KN+P)$} &\multirow{2}{*}{$N$} &\multirow{2}{*}{0} &\multirow{2}{*}{$N$} &\multirow{2}{*}{$K$} &\multirow{2}{*}{${\rm{gcd}}(P,NK)=1$} &{asymptotically}\\
&&&&&&&&{optimal}\\
\cline{2-9}
&\multirow{2}{*}{\emph{Theorem 4}} &\multirow{2}{*}{$p(p-1)$} &\multirow{2}{*}{$p$} &\multirow{2}{*}{$p$} &\multirow{2}{*}{$p-1$} &\multirow{2}{*}{$p$}
&\multirow{2}{*}{$p$ is an odd prime}&{asymptotically}\\
&&&&&&&&{optimal}\\
\hline
\end{tabular}}
\begin{tablenotes}
\item
\fontsize{8pt}{8pt}\selectfont $\theta_{\textrm{max}}$ is the maximum periodic ambiguity magnitude for
$(\tau,v)\in (-Z_x,Z_x)\times (-Z_y,Z_y )$, where $\tau$ is time delay and $v$ is Doppler shift.
\end{tablenotes}
\end{threeparttable}
\end{table*}

In practice, the maximum Doppler shift is often much smaller than the signal bandwidth [15].
Recognizing this, significant efforts have been devoted to minimizing the local ambiguity sidelobes of sequences [15]-[26].
In [16], for example, an energy gradient method was used to optimize the local ambiguity functions of a sequence set.
In [17], a multi-stage accelerated iterative sequential optimization (MS-AISO) algorithm was used to generate sequence sets with enhanced local ambiguity functions in reference to the works in [15], [16].
Although numerous research attempts have been made from the optimization standpoint [15]-[23], only a few works are known on analytical constructions of sequence sets with good local ambiguity functions [14], [24]-[26].
In [14], theoretical bounds on the parameters of unimodular periodic sequence sets with low ambiguity zone (LAZ) and zero ambiguity zone (ZAZ) have been developed.
Meanwhile, based on quadratic phase sequences, a class of unimodular ZAZ sequence sets was introduced in [14]. Doppler-resilient phase-coded waveforms (pulse trains) were designed in [24] by carefully transmitting the two sequences in a Golay pair according to the ``1" and ``0" positions of a binary Prouhet-Thue-Morse (PTM) sequence. Such a construction was then generalized in [25] by applying complete complementary codes and generalized PTM sequences.
Recently, [26] pointed out that a class of binary LCZ sequence sets presented in [4] exhibits low ambiguity properties in a delay-Doppler zone around the origin.

Against the aforementioned background, the main objective of this paper is to look for new analytical constructions of unimodular ZAZ and LAZ sequence sets.
The core idea behind our proposed constructions is motivated by [27], whereby a ZCZ sequence set was generated by modulating a common ``carrier'' sequence with a set of orthogonal ``modulating'' sequences. More constructions on ``modulating" ZCZ sequence sets can be found in [28]-[30].
Nevertheless, the aforementioned works have not looked into the ambiguity functions behavior of these ``modulating" ZCZ sequences. Such a research gap is filled by this work.

Specifically, by looking into the joint impact of delay and Doppler, a generic design of polyphase ZAZ sequence sets is first presented.
Interestingly, such a design also leads to optimal ZCZ sequence sets.
Secondly, we observe from the discrete Fourier transform (DFT) that a Doppler-incurred phase rotation in the time-domain is equivalent to a shift in the frequency-domain.
Thus, it is natural to expect that sequences with comb-like spectrum are resilient to Doppler shifts.
Having this idea in mind, a second construction of polyphase ZAZ sequence sets with comb-like spectrum is developed, where the zero ambiguity sidelobes are guaranteed by their successive nulls in the frequency-domain.
Finally, a connection between polyphase sequence sets and a novel class of mapping functions from $\mathbb{Z}_{p-1}$ to $\mathbb{Z}_p$ is identified, where $p$ is an odd prime.
Such a finding reveals that constructing LAZ sequence sets is equivalent to finding mapping functions that satisfy certain conditions.
By adopting a class of explicit mapping functions, polyphase LAZ sequence sets are derived.
We further show that the proposed ZAZ and LAZ sequence sets are cyclically distinct, thus facilitating their wide use in practical applications.
As a comparison with the known constructions, the parameters of our proposed periodic ZAZ and LAZ sequence sets are listed in Table \uppercase\expandafter{\romannumeral1}.
It is shown that our proposed sequence sets are asymptotically \textit{optimal} with respect to the theoretical bounds in [14].

The remainder of this paper is organized as follows. In Section \uppercase\expandafter{\romannumeral2}, some necessary notations and lemmas are introduced.
In Section \uppercase\expandafter{\romannumeral3}, two constructions of polyphase ZAZ sequence sets are proposed, whereby the spectral characteristics are analyzed for the latter one.
Then, a construction of asymptotically optimal LAZ sequence sets associated with a novel class of mappings is presented in Section \uppercase\expandafter{\romannumeral4}.
Finally, we summarize our work in Section \uppercase\expandafter{\romannumeral5}.

\section{Preliminaries}

In this section, we introduce the definitions of LAZ/ZAZ sequence sets and review the corresponding theoretical bounds.
Besides, the definition of spectral constraints is briefly recalled.
For convenience, we adopt the following notations throughout this paper.
\begin{enumerate}
\item[-] $\mathbb{Z}_L=\left\{0,1,\cdots,L-1\right\}$ is a ring of integers modulo $L$, $\mathbb{Z}^*_L=\mathbb{Z}_{L}\setminus \{0\}$.
\item[-] For a prime $p$, $\mathbb{F}_p=\{0,\alpha^0,\alpha^1,\cdots,\alpha^{p-2}\}$ is the finite field (Galois field $\mathrm{GF}(p)$) with $p$ elements, where $\alpha$ is a primitive element of $\mathbb{F}_p$ with $\alpha^{p-1}=1$.
\item[-] $\omega_{L}={\rm{exp}}\left({2\pi \sqrt{-1}}/{L}\right)$ is a primitive $L$-th complex root of unit.
\item[-] $\langle{t \rangle }_L$ denotes that the integer $t$ is calculated modulo $L$.
\item[-] $\lfloor {c} \rfloor$ denotes the largest integer not greater than $c$.
\item[-] $c^*$ denotes the complex conjugation of a complex value $c$.
\item[-] ${\rm{lcm}}(a, b)$ and ${\rm{gcd}}(a, b)$ denote the least common multiple and the greatest common divisor of positive integers $a$ and $b$, respectively.
\item[-] For positive integers $N$ and $L$, $N|L$ denotes that $N$ is a divisor of $L$.
\item[-] $\bm{a}|| \bm{b}$ denotes the horizontal concatenation of the vectors $\bm{a}$ and $\bm{b}$.
\item[-] $\odot$ denotes the Hadamard product.
\end{enumerate}

\subsection{Ambiguity Functions and Correlation Functions}

We first give the definition of discrete periodic ambiguity function of sequences [9].

\emph{Definition 1:} Let $\bm{a}=\left(a(0),a(1),\cdots, a(L-1)\right)$ and $\bm{b}=\left(b(0),b(1),\cdots, b(L-1)\right)$ be two complex-valued sequences of length $L$.
The periodic ambiguity function of $\bm{a}$ and $\bm{b}$ at time shift $\tau$ and Doppler shift $v$ is given by
\begin{align}
{\rm{AF}}_{\bm{a},\bm{b}}(\tau,v) = \sum_{t = 0}^{N - 1} a(t)\cdot b^*(\langle{t + \tau \rangle }_L) \cdot \omega_L^{vt},
\end{align}
where $ -L<\tau,v< L$.
If $\bm{a}\neq \bm{b}$, ${\rm{AF}}_{\bm{a},\bm{b}}(\tau)$ is called the cross-ambiguity function;
otherwise, it is called the auto-ambiguity function and denoted by ${\rm{AF}}_{\bm{a}}(\tau,v)$.

When the Doppler shift is zero, we have the following definition on periodic correlation functions.

\emph{Definition 2:} Let $\bm{a}=\left(a(0),a(1),\cdots, a(L-1)\right)$ and $\bm{b}=\left(b(0),b(1),\cdots, b(L-1)\right)$ be two complex-valued sequences of length $L$.
The periodic correlation function of $\bm{a}$ and $\bm{b}$ at time shift $\tau$ is defined by
\begin{align}
{\rm{CF}}_{\bm{a},\bm{b}}(\tau) =\sum_{t = 0}^{L - 1} {a(t)\cdot b^*(\langle{t + \tau \rangle}_L)},
\end{align}
where $ -L<\tau< L$.
If $\bm{a}\neq \bm{b}$, ${\rm{CF}}_{\mathbf{a},\mathbf{b}}(\tau)$ is called the  cross-correlation function; otherwise, it is called the auto-correlation function and denoted by ${\rm{CF}}_\mathbf{a} (\tau)$.

Note that when $v = 0$, the ambiguity function ${\rm{AF}}_{\bm{a},\bm{b}}(\tau,0)$ defined in (1) reduces to the correlation function ${\rm{CF}}_{\bm{a},\bm{b}}(\tau)$.

\subsection{Low/Zero Ambiguity Zone (LAZ/ZAZ) Sequences and Zero Correlation Zone (ZCZ) Sequences}

\emph{Definition 3:} Let $\bm{a}=(a(0),a(1),\cdots, a(L-1))$ be a sequence of length $L$.
Consider a delay-Doppler zone $\Pi=(-Z_x,Z_x)\times (-Z_y,Z_y)\subseteq (-L,L)\times (-L,L)$.
The maximum periodic auto-ambiguity sidelobe of $\bm{a}$ over the zone $\Pi$ is defined by
\begin{align}
\theta=\textrm{max}\left\{ \left|{\rm{AF}}_{\bm{a}}(\tau,v)\right|:(0,0)\neq (\tau,v)\in\Pi\right\}.
\end{align}
If $0<\theta\ll L$, $\bm{a}$ is said to be an LAZ sequence and $\Pi$ refers to the low auto-ambiguity zone; if $\theta=0$, $\bm{a}$ is said to be a ZAZ sequence and $\Pi$ the zero auto-ambiguity zone.

\emph{Definition 4:} Let $\mathcal{S}=\left\{\bm{s}_n\right\}_{n=0}^{N-1}$ be a set of $N$ sequences with length $L$.
Consider a delay-Doppler zone $\Pi=(-Z_x,Z_x)\times (-Z_y,Z_y)\subseteq (-L,L)\times (-L,L)$.
The maximum periodic auto-ambiguity sidelobe $\theta _{\rm{A}}$ and the maximum periodic cross-ambiguity magnitude $\theta_{\rm{C}}$ of $\mathcal{S}$ over the zone $\Pi$ are defined by
\begin{align}
\theta_{\rm{A}}=\textrm{max}\bigg\{ \left|{\rm{AF}}_{\bm{s}_n}(\tau,v)\right|:
\left.\begin{array}{ll}
0\leq n\leq N-1,\\
(0,0)\neq (\tau,v)\in\Pi
\end{array}
\right.
\bigg\}
\end{align}
and
\begin{align}
\theta_{{\rm{C}}}=\textrm{max}\bigg\{ \left|{\rm{AF}}_{\bm{s}_n,\bm{s}_{n'}}(\tau,v)\right|:
\left.\begin{array}{ll}
0\leq n\neq n'\leq N-1,\\
(\tau,v)\in\Pi
\end{array}
\right.
\bigg\}
\end{align}
respectively.
Let $\theta _{\textrm{max}}=\textrm{max}\{\theta_{\rm{A}},\theta_{\rm{C}}\}$ be the maximum periodic ambiguity magnitude over the zone $\Pi$.
If {$0<\theta_{\textrm{max}}\ll L$}, $\mathcal{S}$ is referred to as an $\left(L,N,\Pi,\theta _{\textrm{max}} \right)$-LAZ sequence set, where $L$ denotes the sequence length, $N$ the set size, $\Pi$ the low ambiguity zone, and $\theta_{\textrm{max}}$ the maximum periodic ambiguity magnitude over the zone $\Pi$.
If $\theta_{\textrm{max}}=0$, $\mathcal{S}$ is referred to as an
$\left(L,N,\Pi\right)$-ZAZ sequence set.

{\emph{Definition 5:}} Let $\mathcal{S}=\left\{\bm{s}_n\right\}_{n=0}^{N-1}$ be a set of $N$ sequences with length $L$.
If any two sequences $\bm{s}_n$ and $\bm{s}_{n'}$ in $\mathcal{S}$ satisfy the following correlation property,
\begin{align}
{\rm{CF}}_{\bm{s}_n,\bm{s}_{n'}}(\tau)=\left\{
\begin{array}{ll}
L, & n = n',\,\tau=0, \\
0, & n = n',\,0 <|\tau|< Z, \\
0, & n \ne n',\,|\tau|< Z,
\end{array} \right.
\end{align}
where $0\leq n,n'\leq N-1$, $\mathcal{S}$ is referred to as an $(L,N,Z)$-ZCZ sequence set, where $Z$ refers to the ZCZ width.

\subsection{Bounds on LAZ/ZAZ Sequence Sets and ZCZ Sequence Sets}

In [2], Welch derived several correlation lower bounds by evaluating the mini-max value of the inner products of a vector set. Based on the inner product theorem presented in [2], the following lower bounds can be easily obtained for the unimodular periodic LAZ / ZAZ sequence sets and ZCZ sequence sets, as shown in [14] and [31] respectively.

\emph{Lemma 1 ([14]):} For a unimodular $(L,N,\Pi,\theta_{\rm{max}})$-LAZ sequence set with $\Pi =(-Z_x,Z_x)\times (-Z_y,Z_y)$, the maximum periodic ambiguity magnitude satisfies the following lower bound:
\begin{align}
\theta_{\rm{\max}}\ge \frac{L}{\sqrt{Z_y}}\sqrt{\frac{{N{Z_x}{Z_y}}/{L}-1}{N{Z_x}-1}}.
\end{align}

In order to evaluate the closeness between $\theta_{\max}$ and its lower bound, the optimality factor $\rho_{\rm{LAZ}}$ is defined by
\begin{align}
\rho_{\rm{LAZ}}=\frac{\theta _{\rm{\max}}} {\frac{L}{\sqrt{Z_y}}\sqrt{\frac{{N{Z_x}{Z_y}}/{L}-1}{N{Z_x}-1}}}.
\end{align}
In general, $\rho_{\rm{LAZ}}\geq 1$.
If $\rho_{\rm{LAZ}}=1$, the LAZ sequence set is said to be optimal.

By taking ${{\theta }_{\max}}=0$ in \emph{Lemma 1}, we have the following bound on unimodular ZAZ sequence sets.

{\emph{Lemma 2:}} For a unimodular $(L,N,\Pi)$-ZAZ set with $\Pi =(-Z_x,Z_x)\times (-Z_y,Z_y)$, the following upper bound needs to be satisfied:
\begin{align}
NZ_xZ_y\le L.
\end{align}

To analyse the tightness, the zero ambiguity zone ratio ${\rm{ZAZ_{ratio}}}$ is defined by
\begin{align}
{\rm{ZAZ_{ratio}}}=\frac{Z_xZ_y}{L/N}.
\end{align}
In general, ${\rm{ZAZ_{ratio}}}\leq 1$.
If ${\rm{ZAZ_{ratio}}}=1$, the ZAZ sequence set is said to be optimal.

{\emph{Lemma 3 ([31]):}} For an $(L,N,Z)$-ZCZ sequence set, one has
\begin{align}
NZ\leq {L}.
\end{align}
Such a sequence set is called optimal if the above equality holds.

\subsection{Discrete Fourier Transform (DFT) and Spectral-Null Constraints}

{\emph{Definition 6:}} For a time-domain sequence $\bm{a}=(a(0),a(1),\cdots,a(L-1))$ of length $L$, the corresponding frequency-domain dual $\bm{d}=(d(0),d(1),\cdots,d(L-1))$ of length $L$ is defined by taking the $L$-point DFT on $\bm{a}$, i.e.,
\begin{align}
d(i)=\frac{1}{\sqrt{L}} \sum_{t = 0}^{L - 1} a(t)\cdot\omega_L^{-it}, \,0\leq i\leq L-1.
\end{align}

It follows from (12) that the periodic ambiguity function of $\bm{a}$ and $\bm{b}$ at time shift $\tau$ and Doppler shift $v$ in (1) can be represented by
\begin{align}{\rm{AF}}_{\bm{a},\bm{b}}(\tau,v) = \sum_{i = 0}^{L-1} c(i)\cdot d^*(\langle{i + v\rangle}_L) \cdot \omega_L^{i\tau},
\end{align}
where $\bm{c}$ and $\bm{d}$ are the frequency-domain duals corresponding to $\bm{a}$ and $\bm{b}$, respectively.

Consider a wireless system where the entire spectrum is divided into $L$ carriers.
Let us further consider a ``subcarrier marking vector''$\mathbb{c}=[c_0,c_1,\cdots,c_{L-1}]$ with $c_i=1$ if the $i$-th subcarrier is available and $c_i=0$ otherwise.
The ``spectral constraint'' is defined by the set of indices of all forbidden carrier positions, i.e., $\Omega=\left\{i:c_i=0, i\in \mathbb{Z}_L\right\}$.
Suppose multiple terminals or targets are supported with distinct signature sequences over the $L - |\Omega|$ available carriers specified by $\mathbb{Z}_{L}\setminus {\Omega}$ [32].

\emph{Definition 7:} Let $\mathcal{S}=\left\{\bm{s}_n\right\}_{n=0}^{N-1}$ be a set of $N$ sequences with length $L$, $\bm{d}_n=(d_n(0),d_n(1),\cdots,d_n(L-1))$ be the frequency-domain dual corresponding to ${\bm{s}}_n$.
For $\Omega \subset \mathbb{Z}_L$, the sequence set $\mathcal{S}$ is subject to the spectral-null constraint $\Omega$ if
\begin{align}
\sum_{n=0}^{N-1}\left|d_n(i)\right|^2=0
\end{align}
holds for any $i\in \Omega$.

\section{Proposed Constructions of ZAZ Sequence Sets}

{Before the context of the proposed constructions of ZAZ sequence sets, we first review a framework of ZCZ sequence sets from the view point of ``modulating'' [27].}

Let ${\bm{a}}=(a(0),a(1),\cdots,a(MN-1))$ be a sequence of length $MN$ and $\left\{ {\bm{b}_n}\right\}_{n=0}^{N-1} $  a set of $N$ orthogonal sequences with ${\bm{b}}_n=(b_n(0),b_n(1),\cdots,b_n(N-1))$.
By modulating ${\bm{a}}$ with $N$ different orthogonal sequences $\left\{ {\bm{b}_n}\right\}_{n=0}^{N-1}$, a sequence set ${\mathcal{S}} = \left\{ {\bm{s}_n} \right\}_{n=0}^{N-1}$ can be obtained by
\begin{align}
&{{\bm{s}}_n} = {\bm{a}} \odot\big[\underbrace { {{\bm{b}}_n}\,||\,{{\bm{b}}_n}\,|| \,\cdots \,||{\,{\bm{b}}_n}}_{M } \big],
\end{align}
where the $t$-th entry of ${\bm{s}_n}$ with $0\leq t\leq MN-1$ is
\begin{align}
s_n(t) = a(t)\cdot b_n(t\,{\mathrm{mod}}\,N).
\end{align}
The sequence ${\bm{a}}$ can be regarded as a ``carrier'' sequence and ${\bm{b}}_n$ a ``modulating'' sequence.

Inspired by the above framework, by well choosing the carrier sequences, we introduce two constructions of asymptotically optimal unimodular ZAZ sequence sets and show that all the constructed sequences in a ZAZ sequence set are cyclically distinct.

\subsection{The First Proposed Construction of ZAZ Sequence Sets}

\emph{Construction A:}

Consider positive integers $M$, $N$, and $K$ such that $K<N$ and ${\rm{gcd}}(K,N)=1$.
Let $\bm{D}=\left[{d_n(t)}\right]_{n,t=0}^{N-1}$ be an $N\times N$ DFT matrix, where the $t$-th entry of the row  ${\bm{d}}_n$  is $d_n(t)=\omega_N^{Knt}$.
Define a sequence ${\bm{a}}$ of length $MN^2$ by
 \begin{align}
&{\bm{a}} = \notag \\
&\big[ \underbrace { {{\bm{d}}_0}\,||{{\bm{d}}_0}|| \cdots ||{{\bm{d}}_0} }_M||\underbrace { {\bm{d}}_1|| {\bm{d}}_1||\cdots ||{\bm{d}}_1 }_M|| \cdots ||\underbrace {{{\bm{d}}_{N - 1}}|| \cdots ||{{\bm{d}}_{N - 1}}}_M \big],
\end{align}
where the $t$-th entry of ${\bm{a}}$ is $a(t)=\omega_N^{Kt_2t_0}$, $0\leq t\leq MN^2-1$, $t=MNt_2+Nt_1+t_0$, $t_2=\lfloor {t}/(MN) \rfloor$, $t_1 = {\langle\lfloor {t}/{N} \rfloor \rangle}_M$, and $t_0 = {{\langle t \rangle}_N}$.
Following the framework in (15), using the above sequence ${\bm{a}}$ and an orthogonal sequence set $\left\{ {\bm{b}_n}\right\}_{n=0}^{MN-1}$, a sequence set ${\mathcal{S}}=\left\{ {\bm{s}_n} \right\}_{n=0}^{MN-1}$ can be constructed.
Recalling (16), the $t$-th entry of ${\bm{s}}_{n}$ can be expressed as
\begin{align}
s_n(t)=\omega_N^{Kt_2t_0}\cdot b_n(Nt_1+t_0),
\end{align}
where $0\leq t\leq MN^2-1$, $t=MNt_2+Nt_1+t_0$, $t_2=\lfloor {t}/(MN) \rfloor$, $t_1 = {\langle\lfloor {t}/{N} \rfloor \rangle}_M$, and $t_0 = {{\langle t \rangle}_N}$.

{\emph{Theorem 1:}} The sequence set $\mathcal{S}$ constructed above is a unimodular $(MN^2, MN, \Pi)$-ZAZ sequence set with $\Pi =( -\lfloor {N}/{K} \rfloor ,\lfloor {N}/{K} \rfloor )\times ( -K, K )$.

{\emph{Proof:}} We will show that the sequence set $\mathcal{S}$ has ideal ambiguity functions over a delay-Doppler zone around the origin.
Note that for two sequences $\bm{a}$ and $\bm{b}$ of length $L$, the ambiguity function has the symmetry property, i.e.,  ${\rm{AF}}_{{\bm{a}},{\bm{b}}}(\tau,v) = {\rm{AF}}^*_{{\bm{b}},{\bm{a}}}(-\tau,v)$ for $0 \le \tau < L$.
Therefore, in the rest of this paper, we will only discuss the ambiguity function ${\rm{AF}}_{{\bm{a}},{\bm{b}}}(\tau,v)$ with $0 \le \tau< L$ and $ |v| < L$.
Let $\bm{s}_n$ and $\bm{s}_{n'}$ be any two sequences in $\mathcal{S}$, where $0\leq n,n'\leq MN-1$.
Calculate the periodic ambiguity function of $\bm{s}_n$ and $\bm{s}_{n'}$ as follows:
\begin{align}
&{\rm{AF}}_{{\bm{s}_n},{\bm{s}_{n'}}}(\tau,v)\notag\\
=&\sum_{t=0}^{MN^2-1}s_n(t)\cdot s^*_{n'}(\langle{t + \tau \rangle}_{MN^2})\cdot \omega_{MN^2}^{vt} \notag\\
 =& \sum_{t_2=0}^{N - 1} \sum_{t_1 = 0}^{M - 1} \sum_{t_0 = 0}^{N - 1} \omega_N^{Kt_2t_0}\cdot \omega_N^{-K(t_2 + \tau_2 + \delta_{t_1,\tau_1,t_0,\tau_0})(t_0 + \tau_0)}\notag\\
 & \cdot \omega_{MN^2}^{v(M Nt_2 + N{t_1} + {t_0})} \cdot b_n(Nt_1+t_0)\notag\\
 &\cdot b^*_{n'}(\langle N(t_1 + \tau_1 + \delta_{t_0,\tau_0})+(t_0+\tau_0- N\delta_{t_0,\tau_0})\rangle_{MN} ) \notag\\
 =& \sum_{t_1 = 0}^{M - 1}\sum_{t_0 = 0}^{N - 1} \omega_N^{-K(\tau_2 + {\delta_{t_1,\tau_1,t_0,\tau_0}})(t_0+\tau_0)}\cdot \omega_{MN}^{vt_1}\cdot\omega_{MN^2}^{vt_0}\notag\\
 &\cdot b^*_{n'}(\langle N(t_1 + \tau_1 + \delta _{t_0,\tau_0})+(t_0+\tau_0)- N\delta_{t_0,\tau_0}\rangle _{MN})\notag\\
 & \cdot b_n(Nt_1+t_0) \cdot\sum_{t_2 = 0}^{N - 1} \omega_N^{(v - K\tau_0)t_2},
\end{align}
where $t_2=\lfloor {t}/(MN) \rfloor $, $t_1={\langle\lfloor t/N \rfloor \rangle}_M$, $t_0={{\langle t \rangle}_N}$, $n_1=\lfloor {n}/{N} \rfloor $, $n_0={\langle n \rangle}_N$, $\tau=MN\tau_2+N\tau_1+\tau_0$, $\tau_2 = \lfloor {\tau /(MN)} \rfloor $, $\tau_1 = {\langle {\lfloor {\tau/N} \rfloor} \rangle_M}$, $\tau_0 = {\langle \tau \rangle_N}$,
${\delta_{t_0,\tau_0}} = \lfloor {(t_0 + \tau_0) /N}\rfloor $, and
${\delta_{t_1,\tau_1,t_0,\tau_0}} = \lfloor {(t_1 + {\tau_1} + {\delta_{t_0,\tau_0}}) /M}\rfloor $.

Consider the following two cases:

{\emph{Case 1:}} When $v=0$, $\tau =0$, and $n\ne n'$, (19) is simplified to
\begin{align}
{\rm{AF}}_{{\bm{s}_n},{\bm{s}_{n'}}}(0,0)
&= N\cdot\sum_{t_0=0}^{N-1} \sum_{t_1 = 0}^{M - 1} b_n(Nt_1+t_0) \cdot b^*_{n'}(Nt_1+t_0)\notag\\
&=N\cdot\sum_{t = 0}^{MN-1} b_n(t) \cdot b^*_{n'}(t)\notag\\
&=0,
\end{align}
where $\sum_{t = 0}^{MN-1} b_n(t) \cdot b^*_{n'}(t)=0$ for ${n}\neq {n'}$.

{\emph{Case 2:}} When $v=0$ and $0<\tau_0<N $, or $0<|v|< K$ and $0\leq{\tau_{0}}< \lfloor {N}/{K} \rfloor $,
there is $\langle v - K\tau_0 \rangle_N \ne 0$ as ${\rm{gcd}}(K,N)=1$.
Then $\sum_{t_2 = 0}^{N - 1} \omega_N^{(v - K\tau_0)t_2}=0$ holds in (19).
Therefore, we have ${{\rm{AF}}_{{\bm{s}_n}, {\bm{s}_{n'}}}}(\tau,v)=0$.

From the above discussions, we assert that when $|\tau|<\left\lfloor {N}/{K} \right\rfloor$ and $|v|< K$, the auto-ambiguity function ${\rm{AF}}_{{\bm{s}_n}}(\tau,v)=0$ for $(\tau,v)\neq (0,0)$ and the cross-ambiguity function ${\rm{AF}}_{{\bm{s}_n},{\bm{s}_{n'}}}(\tau,v)=0$ for $n\neq n'$.
Consequently, the sequence set $\mathcal{S}$ has ideal ambiguity properties over the delay-Doppler zone $(-\lfloor {N}/{K}\rfloor ,\lfloor {N}/{K} \rfloor )\times (-K,K)$.

Note that cyclically equivalent sequences are not treated as essentially different sequences and thus are not desired in practical applications [1].
In the following, a specific orthogonal sequence set $\left\{ {\bm{b}_n}\right\}_{n=0}^{MN-1}$ is provided to guarantee that all the sequences in the set $\mathcal{S}$ derived from  \emph{Construction A} are cyclically distinct.

Let $\bm{C}=\left[c_{i}(j)\right]_{i,j=0}^{M-1}$ be an $M\times M$ DFT matrix with $c_{i}(j)=\omega_M^{i j}$ and $\bm{D} =\left[d_{i}(j)\right]_{i,j=0}^{N-1}$ a generalized $N\times N$ DFT matrix with $d_{i}(j)=\omega_N^{i \sigma(j)}$, where $\sigma$ is a permutation of  $\mathbb{Z}_N$ such that $\sigma(j) \neq \alpha j+\beta$ for any $\alpha,\beta\in\mathbb{Z}_N$.
Define the orthogonal matrix $\bm{B}=\left[b_{n}(t)\right]_{n,t=0}^{MN-1}$ as the Kronecker product of $\bm{C}$ and $\bm{D}$, where the $t$-th entry of the row $\bm{b}_{n}$ is $b_{n}(t)=\omega_M^{n_1t_1} \cdot\omega_N ^ {n_0 \sigma (t_0)}$, $n_1=\lfloor {n}/{N} \rfloor $, $n_0={\langle n \rangle}_N$, $t_1=\lfloor {t}/{N} \rfloor $, and $t_0={\langle t \rangle}_N$.
Then, using the orthogonal sequence set $\left\{ {\bm{b}_n}\right\}_{n=0}^{MN-1}$, a sequence set ${\mathcal{S}}=\left\{ {\bm{s}_n} \right\}_{n=0}^{MN-1}$ can be constructed following (18) in \emph{Construction A}.
The $t$-th entry of ${\bm{s}}_{n}$ is given by
\begin{align}
s_n(t)=\omega_N^{Kt_2 t_0+n_0 \sigma(t_0)}\cdot \omega_M^{n_1 t_1},
\end{align}
where $0\leq t\leq MN^2-1$, $t=MNt_2+Nt_1+t_0$, $t_2=\lfloor {t}/(MN) \rfloor$, $t_1={\langle\lfloor {t}/{N} \rfloor \rangle}_M$, $t_0={{\langle t \rangle}_N}$, $n=Nn_1+n_0$, $n_1=\lfloor {n}/{N} \rfloor $, and $n_0={\langle n \rangle}_N$.

\emph{Corollary 1:} The sequence set $\mathcal{S}$ constructed from (21) is a polyphase  $(MN^2, MN, \Pi)$-ZAZ sequence set with $\Pi =(-\lfloor {N}/{K} \rfloor, \lfloor {N}/{K} \rfloor )\times (-K,K)$.
All the sequences in $\mathcal{S}$ are cyclically distinct.

{\emph{Proof:}} It follows directly from \emph{Theorem 1} that $\mathcal{S}$ is a polyphase $(MN^2,MN,\Pi)$-ZAZ sequence set with $\Pi =(-\lfloor {N}/{K} \rfloor ,\lfloor {N}/{K} \rfloor)\times (-K,K)$.
Next, we will show that all the sequences in $\mathcal{S}$ are cyclically distinct.
Assume on the contrary that $\bm{s}_n$ and $\bm{s}_{n'}$ with $0\leq n\neq n'\leq MN-1$ in $\mathcal{S}$ are cyclically equivalent at the time shift $\tau$.
It implies that
\begin{align}
s_n(t)= s_{n'}(\langle{t + \tau \rangle }_{MN^2})\cdot \omega_{MN^2}^c
\end{align}
holds for all $0\leq t \leq MN^2-1$, where $c \in\mathbb{Z}_{MN^2}$.
It follows from (21) that for all $0\leq t_2 \leq N-1$, $0\leq t_1 \leq M-1$, and $0\leq t_0 \leq N-1$, there is
\begin{align}
&\omega_N^{{n_0}\sigma (t_0) - n'_0\sigma (t_0 + \tau_0 - N\delta_{t_0,\tau_0})}\cdot \omega_N^{-K(\tau_2 + \delta_{t_1,\tau_1,t_0,\tau_0})(t_0 + \tau_0)} \notag\\
&\cdot \omega_M^{ -n'_1(\tau_1 + \delta _{t_0,\tau_0})}\cdot\omega_M^{(n_1 - n'_1)t_1} \cdot \omega_N^{-K\tau_0t_2} = \omega_{MN^2}^c.
\end{align}
Note that for any $0\leq t_2\leq N-1$, (23) holds if and only if $\tau_0 = 0$.
Then, we have $\delta_{t_0,\tau_0} = 0$, and (23) becomes
\begin{align}
&\omega_N^{(n_0 - n'_0)\sigma (t_0)} \cdot \omega _N^{-K(\tau_2 + \delta_{t_1,\tau_1,t_0,0})t_0} \cdot \omega _M^{-n'_1\tau_1}\cdot \omega_M^{(n_1 - n'_1)t_1} \notag\\
&=\omega_{MN^2}^c.
\end{align}
For $0 \le t_1 < M - {\tau_1}$ and $\delta_{t_1,\tau_1,t_0,0}=0$, or $M - \tau_1 \le t_1 < M$ and $\delta_{t_1,\tau_1,t_0,0}=1$,
(24) holds if and only if $n_1 = n'_1$. Thus, (24) simplifies to
\begin{align}
\omega_N^{(n_0 - n'_0)\sigma (t_0)} \cdot \omega_N^{-K(\tau_2 + \delta _{t_1,\tau_1,t_0,0})t_0}\cdot \omega_M^{-n'_1\tau_1} = \omega_{MN^2}^{c}.
\end{align}
For $0 \le t_1 < M $, (25) holds if and only if $\delta_{t_1,\tau_1,t_0,0}=0$.
Then, we have $\tau_1=0$, and (25) becomes
\begin{align}
\omega_N^{(n_0 - n'_0)\sigma (t_0)} \cdot \omega_N^{-K\tau_2 t_0}
= \omega_{MN^2}^c.
\end{align}
Since $n\neq n'$ and $n_1=n'_1$, there is $ n_0\neq n'_0$.
Then, it follows from the above equation that $\sigma (t_0) \equiv \frac{K\tau_2}{n'_0-n_0}t_0 + y\bmod N$ for all $0\leq t_0 \leq N-1$, where $y\in\mathbb{Z}_N$.
Obviously, it is impossible since $\sigma(t_0) \neq xt_0 +y$ for any $x,y\in\mathbb{Z}_N$.
Consequently, we deduce that all the sequences in $\mathcal{S}$ are cyclically distinct.

{\begin{table*}[t]
\centering
{\caption{{Comparison of some known optimal polyphase ZCZ sequence sets}}
{\begin{tabular}{|c|c|c|c|c|c|}
\hline
Method &Length &Set size &ZCZ width &Alphabet size &Constraint \\
\hline
{[28]} &$2^{n+2}$ &$2^n$ &$4$ &$4$ &\\
\hline
{[29]}
&$N^2$ &$N$ &$N$ &$N$ & $N$ is an odd prime\\
\hline
\multirow{4}{*}{[34]}
&\multirow{2}{*}{$\prod_{k=0}^{n-1}M_k=L^2$} &\multirow{2}{*}{$\prod_{k=0}^{n-p-1}M_k$} & \multirow{2}{*}{$\prod_{k=n-p}^{n-1}M_k$} & \multirow{2}{*}{$L^2$} &$0<p\leq n-1$, positive integers $M_k$ with\\
&&&&& $0\leq k \leq n-1$ are not necessarily distinct\\
\cline{2-6}
&\multirow{2}{*}{$MN$} &\multirow{2}{*}{$N$} &\multirow{2}{*}{$M$} &\multirow{2}{*}{${\rm{lcm}}(N,r)$} &${\rm{gcd}}(M,N)=1$, $r$ is the alphabet size\\
&&&&& of a perfect sequence with length $M$ \\
\hline
{[27]} &$MN$ &$N$ &$M$ &$MN$ &\\
\hline
\multirow{2}{*}{[33]} &$MN$   &$N$  &$M$ &not less than $MN$&\\
\cline{2-6}
        &$M^2N$ &$N$  &$M^2$ &$MN$&\\
\hline
{[30]} &$MN^2$ &$N$ &$MN$ &$MN$ & $M$ is a square-free integer\\
\hline
\emph{Corollary 2} &$MN^2$ &$MN$ &$N$ &${\rm{lcm}}(M,N)$ & \\
\hline
\end{tabular}}
}\end{table*}}

\begin{figure*}[!t]
 \centering
\subfigure[{The auto-ambiguity magnitudes of ${\bm{s}}_0$ over $[-8,8]\times [-8,8]$.
 }]{
 \includegraphics[width=0.315\textwidth]{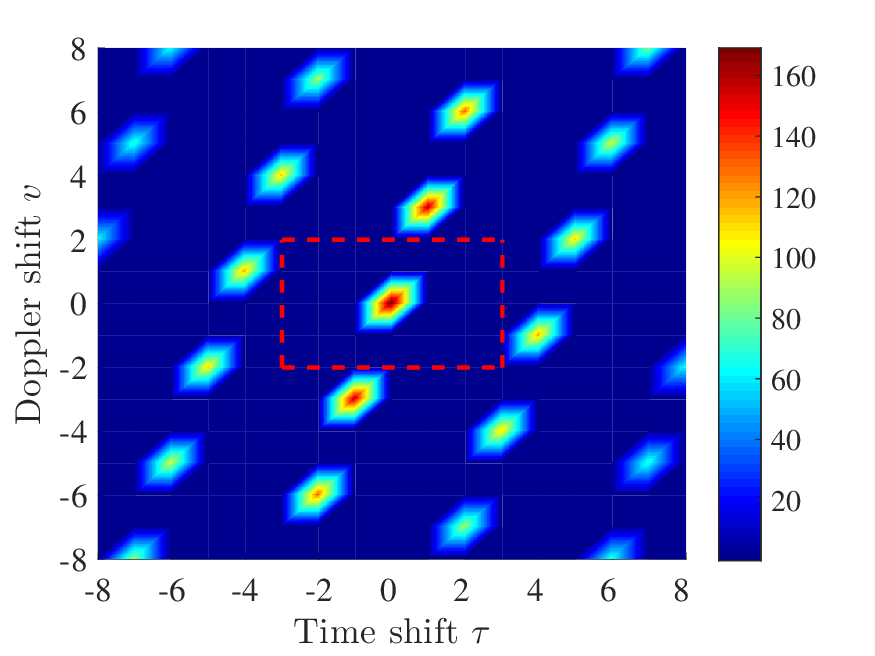}}
\subfigure[{The auto-ambiguity magnitudes of ${\bm{s}}_0$ over $[-3,3]\times [-2,2]$.}
]{
 \includegraphics[width=0.315\textwidth]{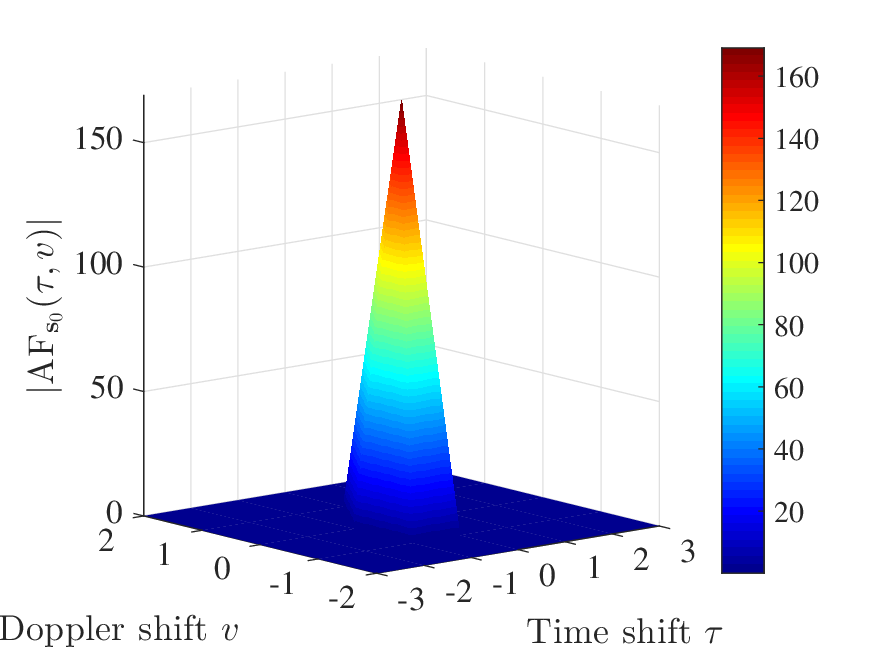}
 }
 \subfigure[{The cross-ambiguity magnitudes of ${\bm{s}}_0$ and ${\bm{s}_1}$ over $[-8,8]\times [-8,8]$.}
 ]{
 \includegraphics[width=0.315\textwidth]{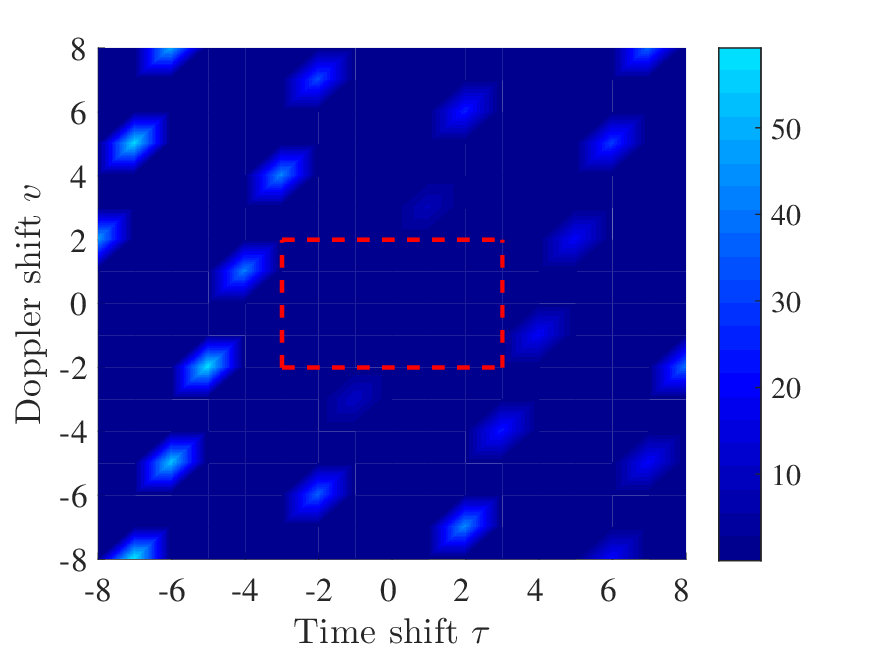}
 }
\caption{The ambiguity magnitudes of ${\bm{s}}_0$ and ${\bm{s}_1}$ in $\mathcal{S}$ from \emph{Example 1}.}
\end{figure*}

{\emph{Remark 1:}} In \emph{Corollary 1}, all the sequences in $\mathcal{S}$ are cyclically distinct if the permutation $\sigma$ of $\mathbb{Z}_N$ satisfies $\sigma(j) \neq xj +y$ for any $x,y\in\mathbb{Z}_N$.
For example, for an odd prime $N$ and an integer $\alpha$ with $1<\alpha<N$, the permutation $\sigma(j) = \langle{j^\alpha\rangle}_N$ satisfies this condition if ${\rm{gcd}}(N-1,\alpha) = 1$.
For a fixed $N$, the zero ambiguity zone $\Pi =(-\lfloor {N}/{K}\rfloor ,\lfloor {N}/{K} \rfloor)\times (-K,K)$ of the proposed $(MN^2,MN,\Pi)$-ZAZ sequence set ${\mathcal{S}}$ can be set flexibly by changing $K$.
According to (10), the zero ambiguity zone ratio of $\mathcal{S}$ is
\begin{align}
{\rm{ZAZ_{ratio}}}=\frac{K}{N} \cdot\left\lfloor {\frac{N}{K}}\right\rfloor.
\end{align}
Note that $\mathop {\lim}\limits_{\langle N\rangle_K \rightarrow 0}{\rm{ZAZ_{ratio}}}\rightarrow 1$, implying that the constructed ZAZ sequence set $\mathcal{S}$ is asymptotically optimal with respect to the theoretical bound in \emph{Lemma 2}.

{\emph{Corollary 2:}} When $K=1$, the sequence set ${\mathcal{S}}$ derived from (21) is an optimal polyphase $\left(MN^2,MN,N \right)$-ZCZ sequence set.
All the sequences in ${\mathcal{S}}$ are cyclically distinct.

{\emph{Proof:}} It follows directly from \emph{Corollary 1} that when $K=1$, ${\mathcal{S}}$ is an $\left( MN^2,MN,N \right)$-ZCZ sequence set and all the sequences are cyclically distinct.
The parameters of ${\mathcal{S}}$ achieve the theoretical bound in \emph{Lemma 3}, and therefore ${\mathcal{S}}$ is optimal.

\emph{Remark 2:} Due to their important applications in quasi-synchronous code-division multiple-access (QS-CDMA) systems, ZCZ sequences have been extensively investigated [27]-[30], [33], [34].
As a comparison, the parameters of some known optimal polyphase ZCZ sequence sets are listed in Table \uppercase\expandafter{\romannumeral2}.
In [29], a construction of $\left(N^2, N, N \right)$-ZCZ sequence sets based on perfect nonlinear functions (PNFs) was proposed.
When $M=1$, the ZCZ sequence set ${\mathcal{S}}$ in {\emph{Corollary 2}} simplifies to that in [29], where the ``carrier'' sequence $\bm{a}$ of length $N^2$ is defined by $a(t) = \omega_N ^{\lfloor {t}/N \rfloor t}$.
When $M>1$, however, the $\left( MN^2, MN, N \right)$-ZCZ sequence set ${\mathcal{S}}$ in \emph{Corollary 2} is new, in which the ``carrier'' sequence $\bm{a}$ of length $MN^2$ with $a(t) = \omega_N ^{\lfloor {t}/(MN) \rfloor t} $ is an extension of the perfect sequence by generalizing the PNF in [29].

Here, we give an example to illustrate the proposed construction.

{\emph{Example 1:}} Let $M=1$, $N=13$, $K=3$, and the permutation $\sigma(j)={{\left\langle j^5 \right\rangle}_{13}}$ for $ j\in \mathbb{Z}_{13}$.
Following (21), a polyphase $\left(169,13,\Pi \right)$-ZAZ sequence set ${\mathcal{S}}=\left\{ {\bm{s}_n}\right\}_{n=0}^{12}$ with $\Pi =(-4,4)\times (-3,3)$ can be derived, where the $t$-th entry of ${\bm{s}}_{n}$ is given by
\begin{align*}
s_n(t)=\omega _{13}^{3t_2 t_0+n \sigma(t_0)},
\end{align*}
$0\leq t\leq 168$, $t_2=\lfloor {t}/13 \rfloor$, and $t_0={\langle t \rangle}_{13}$.
The zero ambiguity zone ratio of ${\mathcal{S}}$ is ${\rm{ZAZ_{ratio}}}=0.923077$.
The auto-ambiguity magnitudes of the sequence ${\bm{s}}_{0}$ over $[-8,8]\times [-8,8]$ and $[-3,3]\times [-2,2]$, and the cross-ambiguity magnitudes of the sequences ${\bm{s}}_{0}$ and ${\bm{s}}_{1}$ over $[-8,8]\times [-8,8]$ are shown in Fig. 1 (a), Fig. 1 (b), and Fig. 1 (c) respectively.
It can be seen that the sequence ${\bm{s}}_0$ has zero auto-ambiguity sidelobes over $[-3,3]\times [ -2,2]$, exhibiting a thumbtack shape over this zone, ${\bm{s}}_0$ and ${\bm{s}}_1$ have zero cross-ambiguity magnitudes over $[-3,3]\times [-2,2]$.

\subsection{The Second Proposed Construction of ZAZ Sequence Sets}

{Note that according to the DFT, a Doppler-incurred phase rotation in the time-domain corresponds to a shift in the frequency-domain.
Therefore, zero ambiguity functions for local non-zero Doppler shifts can be guaranteed if each non-zero element of the frequency-domain duals is followed by successive nulls according to (13).
With this idea, we have the following theorem. }

{\emph{Theorem 2:}} Consider a unimodular $\left(L,N,Z \right)$-ZCZ sequence set subject to the spectral-null constraint $\Omega$. For any $i \in\mathbb{Z}_L \setminus {\Omega}$ and $0< |v|< K$, if $\langle{i+v\rangle}_{L} \in \Omega$, then $\mathcal{S}$ is a unimodular $\left(L,N,\Pi \right)$-ZAZ sequence set with $\Pi =(-Z,Z) \times (-K,K)$.

{{\emph{Proof:}} For the $\left(L,N,Z \right)$-ZCZ sequence set $\mathcal{S} = \left\{{\bm{s}_n}\right\}_{n=0}^{N-1}$, let $\bm{s}_n$ and $\bm{s}_{n'}$ be any two sequences in $\mathcal{S}$, where $0\leq n,n'\leq N-1$.
Consider the periodic ambiguity function ${\rm{AF}}_{{\bm{s}_n},{\bm{s}_{n'}}}(\tau,v)$ of $\bm{s}_n$ and $\bm{s}_{n'}$ in the following two cases:}

{{\emph{Case 1:}} When $v=0$, we have ${{\rm{AF}}_{\bm{s}_n}}(\tau,0)={{\rm{CF}}_{\bm{s}_n}}(\tau)=0$ for $0< \tau < Z$, and
${{\rm{AF}}_{{\bm{s}_n},{\bm{s}_{n'}}}}(\tau,0)={{\rm{CF}}_{{\bm{s}_n},{\bm{s}_{n'}}}}(\tau)=0$  for $n\ne n'$ and $0 \leq \tau < Z$ according to the ZCZ property of $\mathcal{S}$.}

{{\emph{Case 2:}} When $0< |v|< K$, according to (13), the periodic ambiguity function of $\bm{s}_n$ and $\bm{s}_{n'}$ can be represented by
\begin{align}
{\rm{AF}}_{{\bm{s}_n},{\bm{s}_{n'}}}(\tau,v)=&\sum_{i\in {\Omega}}d_n(i)\cdot d^*_{n'}(\langle{i + v \rangle}_{L}) \cdot \omega _{L}^{\tau i}\notag\\
&+\sum_{i\in \mathbb{Z}_L \setminus {\Omega}}d_n(i)\cdot d^*_{n'}(\langle{i + v \rangle}_{L}) \cdot \omega _{L}^{\tau i},
\end{align}
where $\bm{d}_n$ and $\bm{d}_{n'}$ are the frequency-domain duals corresponding to the sequences $\bm{s}_n$ and $\bm{s}_{n'}$, respectively.
Note that when $i\in \Omega$, ${d_n}(i)=0$ holds in (28).
When $i\in \mathbb{Z}_{L} \setminus {\Omega}$, there is $\langle{i + v \rangle}_{L}\in \Omega$ as $0< |v|< K$, then $d^*_{n'}(\langle{i + v \rangle}_{L})=0$ holds in (28).
Therefore, we have ${\rm{AF}}_{{\bm{s}_n},{\bm{s}_{n'}}}(\tau,v)=0$ for $0< |v|< K$.}

Combining the above two cases, we assert that when $|\tau|<Z$ and $|v|< K$, the auto-ambiguity function ${\rm{AF}}_{{\bm{s}_n}}(\tau,v)=0$ for $(\tau,v)\neq (0,0)$ and the cross-ambiguity function ${\rm{AF}}_{{\bm{s}_n},{\bm{s}_{n'}}}(\tau,v)=0$ for $n\neq n'$.
Therefore, the sequence set $\mathcal{S}$ has ideal ambiguity properties over the delay-Doppler zone $(-Z,Z)\times (-K,K)$.

In the following, based on ZCZ sequence sets, a simple construction of ZAZ sequence sets is proposed by imposing comb-like spectrum.

\emph{Corollary 3:} For an optimal unimodular $\left(L,N,Z \right)$-ZCZ sequence set $\mathcal{A}$, by duplicating each sequence $K$ times, an optimal unimodular $\left( KL,N,\Pi \right)$-ZAZ sequence set ${\mathcal{S}}$ with $\Pi =(-Z,Z) \times (-K,K)$ is obtained.

{\emph{Proof:}} It is easy to verify that ${\mathcal{S}}$ is a $\left(KL,N,Z \right)$-ZCZ sequence set.
By duplicating each sequence of length $L$ in $\mathcal{A}$ $K$ times, $K$ successive nulls are uniformly distributed in the frequency-domain duals corresponding to the sequences of length $KL$ in ${\mathcal{S}}$, i.e., ${\mathcal{S}}$ is subject to the spectral-null constraint $\Omega= \left\{ K\alpha+\beta: \alpha \in \mathbb{Z}_L, \, \beta\in \mathbb{Z}^*_K\right\}$.
Note that for any $i \in\mathbb{Z}_{K L} \setminus {\Omega}= \left\{ K\alpha: \alpha \in \mathbb{Z}_L \right\}$, there is $\langle{i+v\rangle}_{L} \in \Omega$ for $0< |v|< K$.
Then, it follows directly from \emph{Theorem 2} that $\mathcal{S}$ is an optimal unimodular $\left( KL,N,\Pi \right)$-ZAZ sequence set with $\Pi =(-Z,Z) \times (-K,K)$.
According to \emph{Lemma 2}, the parameters of $\mathcal{S}$ achieve the theoretical bound, and therefore $\mathcal{S}$ is optimal.

\emph{Corollary 3} presents a construction of optimal ZAZ sequence sets based on ZCZ sequence sets.
However, such a construction is trivial.
In the sequel, a novel construction of non-trivial ZAZ sequence sets with comb-like spectrum is proposed.

\emph{Construction B:}

Let $K$, $N$, and $P$ be positive integers with $P<K$.
Consider a $(KN+P)\times(KN+P)$ DFT matrix $\bm{D}=\left[d_{i}(j)\right]_{i,j=0}^{KN+P-1}$
with $d_{i}(j)=\omega_{KN+P}^{i j}$.
By selecting $N$ columns from $\bm{D}$ at intervals of $K$, we can obtain a $(KN+P)\times N$ matrix, i.e.,
\begin{align}
&\bm{A} = \notag\\
&{\footnotesize{\left[\setlength{\arraycolsep}{1.2pt}{\begin{array}{cccc}
d_{0}(K\cdot0)&d_{0}(K\cdot 1)& \cdots &d_{0}(K\cdot (N-1))\\
d_{1}(K\cdot0)&d_{1}(K\cdot 1)& \cdots &d_{1}(K\cdot (N-1))\\
 \vdots & \vdots  & \ddots &\vdots \\
d_{KN + P-1}(K\cdot0)&d_{KN + P-1}(K\cdot 1)& \cdots &d_{KN + P-1}(K\cdot (N-1))
\end{array}} \right]}}.
\end{align}
By concatenating the successive rows of $\bm{A}$, a sequence ${\bm{a}}$ of length $N(KN+P)$ is obtained, where the $t$-th entry of ${\bm{a}}$ is $a(t)=\omega_{KN+P}^{Kt_1t_0}$, $0\leq t\leq N(KN+P)-1$, $t=Nt_1+t_0$, $t_1=\lfloor t/N \rfloor$, and $t_0={\langle t \rangle}_N$.
Consider an orthogonal sequence set $\left\{{\bm{b}_n}\right\}_{n=0}^{N-1}$ with ${b}_n(t)=\omega_N^{nt}$, $0\leq t\leq N-1$.
Following the framework in (15), a sequence set ${\mathcal{S}}=\left\{ {\bm{s}_n} \right\}_{n=0}^{N-1}$ can be constructed.
The $t$-th entry of ${\bm{s}}_n$ is given by
\begin{align}
s_n(t)=\omega_{KN+P}^{Kt_1t_0}\cdot \omega_N^{nt_0},
\end{align}
where $0\leq t\leq N(KN+P)-1$, $t=Nt_1+t_0$, $t_1=\lfloor t/N \rfloor$, and $t_0={\langle t \rangle}_N$.

{\emph{Theorem 3:}} The sequence set $\mathcal{S}$ constructed above has the following properties:
\begin{enumerate}
\item It is a polyphase $\left(N(KN+P),N,\Pi \right)$-ZAZ sequence set with $\Pi =(-N,N)\times (-K,K)$;
\item It is subject to the spectral-null constraint $\Omega=\left\{(KN+P)\alpha+K\beta+\gamma: \alpha,\beta\in\mathbb{Z}_N,\,\gamma\in\mathbb{Z}^*_K\right\}\cup\left\{ K N +(KN+P)\alpha+\beta:\alpha\in\mathbb{Z}_N,\,\beta\in\mathbb{Z}_P\right\}$;
\item All the sequences in $\mathcal{S}$ are cyclically distinct.
\end{enumerate}

{\emph{Proof:}} 1) We first show that $\mathcal{S}$ is a ZCZ sequence set.
Let $\bm{s}_n$ and $\bm{s}_{n'}$ be any two sequences in $\mathcal{S}$, where $0\leq n,n'\leq N-1$.
The periodic correlation function of $\bm{s}_n$ and $\bm{s}_{n'}$ is
\begin{align}
 &{\rm{CF}}_{{\bm{s}_n},{\bm{s}_{n'}}}(\tau)
 \notag\\
=&\sum_{t=0}^{N(KN+P)-1}{s_n}(t)\cdot s^*_{n'}(\langle{t+\tau \rangle}_{N(KN+P)} ) \notag\\
=&\sum_{t_1=0}^{ KN+P-1}\sum_{t_0=0}^{N-1}\omega _{KN+P}^{Kt_1t_0}\cdot \omega_{KN+P}^{-K(t_1+\tau_1+\delta_{t_0,\tau_0})(t_0+\tau_0-N\delta_{t_0,\tau_0})}\notag\\
&\cdot \omega_N^{nt_0}\cdot \omega_N^{-n'(t_0+\tau_0)} \notag\\
=&\omega _N^{-n'\tau_0}\cdot \sum_{t_0=0}^{N-1}\omega_{KN+P}^{-K\cdot(\tau_1+\delta_{t_0,\tau_0})(t_0+ \tau_0-N\delta_{t_0,\tau_0})}\cdot\omega_N^{(n-n')t_0}\notag\\
&\cdot \sum_{t_1=0}^{KN+P-1}{\omega_{KN+P}^{-K(\tau_0-N\delta_{t_0,\tau_0} )t_1}},
\end{align}
where $t_1=\lfloor {t}/{N} \rfloor $, $t_0={\langle t \rangle }_N$, $\tau=N\tau_1+\tau_0$, $\tau_1=\lfloor \tau /N \rfloor $, $\tau_0={\langle \tau \rangle}_N$, and $\delta_{t_0,\tau_0} = \lfloor (t_0+\tau_0)/N \rfloor$.

{\emph{Case 1:}} When $\tau=0$ and $n\ne n'$, we have
\begin{align}
{\rm{CF}}_{\bm{s}_n,\bm{s}_{n'}}(0)&=(KN+P)\cdot\sum_{t_0=0}^{N-1}\omega_N^{(n-n')t_0}=0,
\end{align}
where $0<|n-n'|\leq N-1$.

{\emph{Case 2:}} When $0<\tau_0< N$, we have $K(\tau_0-N\delta_{t_0,\tau_0})\not\equiv 0\,{\rm{mod}}\,(KN+P)$.
Then $\sum_{t_1=0}^{KN+P-1}{\omega_{KN+P}^{-K (\tau_0-N\delta_{t_0,\tau_0})t_1}}=0$ holds in (31), implying that ${\rm{CF}}_{{\bm{s}_n},{\bm{s}_{n'}}}(\tau)=0$.

Combining the above two cases, when $0\leq\tau< N$, the cross-correlation function ${\rm{CF}}_{{\bm{s}_n},{\bm{s}_{n'}}}(\tau)=0$ for $n\ne n'$ and the auto-correlation function ${\rm{CF}}_{{\bm{s}_n}}(\tau)=0$  for $\tau\ne 0$.
Consequently, the sequence set $\mathcal{S}$ is an $\left(N(KN+P),N,N \right)$-ZCZ sequence set.

2) Next, we discuss the frequency-domain duals corresponding to the sequences in $\mathcal{S}$.
Let $\bm{d}_n=\left(d_n(0),d_n(1),\cdots,d_n(N(KN+P)-1)\right)$ be the frequency-domain dual corresponding to the sequence $\bm{s}_n$ in $\mathcal{S}$, where $0\leq n\leq N-1$.
We have
\begin{align}
&\sqrt {N(KN+P)} d_n(i)
= \sum_{t= 0}^{N(KN+P)-1} s_m (t) \cdot \omega_{N(KN +P)}^{-it} \notag\\
&= \sum_{t_0=0}^{N-1} {\omega_N^{nt_0} \cdot \omega_{N(KN+P)}^{-it_0}} \cdot \sum_{t_1=0}^{KN+P-1}\omega_{KN+P}^{(Kt_0-i)t_1},
\end{align}
where $t_1=\lfloor t/N \rfloor$ and $t_0={\langle t \rangle }_N$.
Note that when $Kt_0-i\not \equiv0\,{\rm{mod}}\, (KN+P)$, i.e., $i\in \Omega$, where $\Omega= \left\{ (KN+P)\alpha + K\beta+\gamma: \alpha, \beta \in \mathbb{Z}_N,\, \gamma \in \mathbb{Z}^*_K\right\}\cup\left\{K N+(KN+P)\alpha + \beta: \alpha \in \mathbb{Z}_N,\, \beta \in \mathbb{Z}_P \right\}$, there is $\omega _{KN + P}^{(Kt_0 - i)t_1}=0$ in (33), then ${d_n}(i)=0$.
Otherwise, when $i\in \mathbb{Z}_{N(KN+P)} \setminus {\Omega} =\{(KN+P)\alpha+K\beta: \alpha,\beta\in\mathbb{Z}_N\}$, there exists only one solution $t'_0$ with $0\leq t'_0\leq N-1$ such that $K t'_0- i\equiv0\,{\rm{mod}}\,(KN+P)$,
then $\left|d_n(i)\right| =\frac{1}{\sqrt {N(KN+P)}} \left|(KN+P) \cdot{\omega_N^{nt'_0} \cdot \omega_{N( KN +P)}^{-it'_0}} \right| = \sqrt{K+\frac{P}{N}}$.
Therefore, for any $i\in \Omega$, we have $\sum_{n=0}^{N-1}\left|d_n(i)\right|^2=0$.

Note that the sequence set $\mathcal{S}$ is subject to the spectral-null constraint $\Omega$, and for any $i \in\mathbb{Z}_{N(KN+P)} \setminus {\Omega} $, $\langle{i+v\rangle}_{N(KN+P} \in \Omega$ holds for $0< |v|< K$.
Therefore, it follows directly from \emph{Theorem 2} that the $\left(N(KN+P),N,N \right)$-ZCZ sequence set $\mathcal{S}$ is a unimodular $\left( N(KN+P),N,\Pi \right)$-ZAZ sequence set with $\Pi =(-N,N) \times (-K,K)$.

3) Here, we prove that all the sequences in $\mathcal{S}$ are cyclically distinct.
Assume on the contrary that $\bm{s}_n$ and $\bm{s}_{n'}$ with $0\leq n\neq n'\leq N-1$ are cyclically equivalent at the time shift $\tau$. Then
\begin{align}
s_n(t) = s_{n'}(\langle{t + \tau \rangle}_{N(KN+P)}) \cdot \omega_{N(KN+P)}^c
\end{align}
holds for all $0\leq t\leq N(KN+P)-1$, where $c\in \mathbb{Z}_{N(KN+P)}$.
It follows from (30) that for all $0\leq t_1\leq KN+P-1$ and $0\leq t_0\leq N-1$, there is
\begin{align}
&\omega _N^{n'\tau_0}\cdot \omega_N^{(n'-n)t_0} \cdot \omega_{KN +P}^{K(\tau_1 + {\delta_{t_0,\tau_0}})(t_0+\tau_0 - \delta_{t_0,\tau_0}N)} \notag\\
&\cdot\omega_{KN +P}^{K(\tau_0-N\delta_{t_0,\tau_0})t_1}= \omega_{N(KN+P)}^{-c}.
\end{align}
Note that (35) holds for $0\leq t_1\leq KN+P-1$ if and only if $\tau_0 - N{\delta_{t_0,\tau_0}}=0$.
Then, we have $\tau_0 = 0$ and $\delta_{t_0,\tau_0}=0$, and (35) simplifies to
\begin{align}
\omega_{N(KN+P)}^{((KN+P)(n'-n) +K N\tau_1)t_0} = \omega_{N(KN +P)}^{-c}.
\end{align}
This equation holds for $0\leq t_0\leq N-1$ if and only if $(KN+P)(n'-n) +K N\tau_1\equiv0\,{\rm{mod}}\,N(KN +P)$.
It implies that $n=n'$ and $\tau_1=0$ since $P<K$, which contradicts with the condition that $n \neq n'$.
Therefore, we assert that all the sequences in $\mathcal{S}$ are cyclically distinct.

\begin{figure*}[!t]
\centering{\includegraphics[width=1\textwidth]{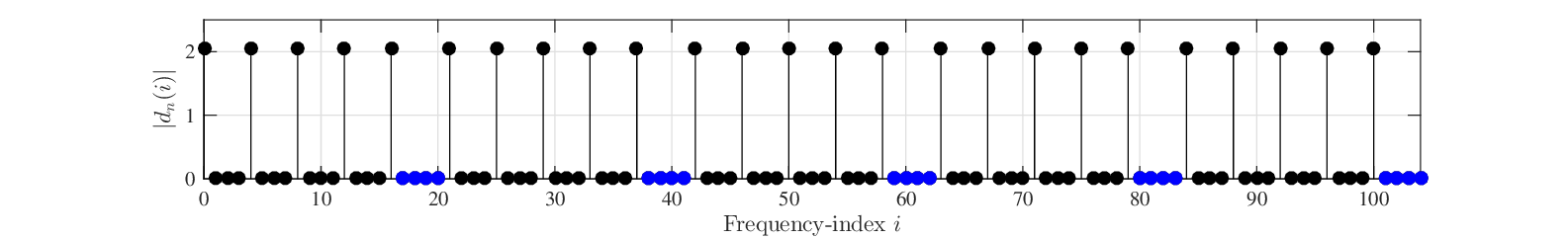}}
\caption{The magnitudes of the frequency-domain dual ${\bm{d}_n}$ corresponding to ${\bm{s}_n}$ in \emph{Example 2}, $0\leq n\leq 4$.}
\end{figure*}
\begin{figure*}[!t]
 \centering
\subfigure[{The auto-ambiguity magnitudes of ${\bm{s}}_0$ over $[-8,8]\times [-8,8]$.
 }]{
 \includegraphics[width=0.315\textwidth]{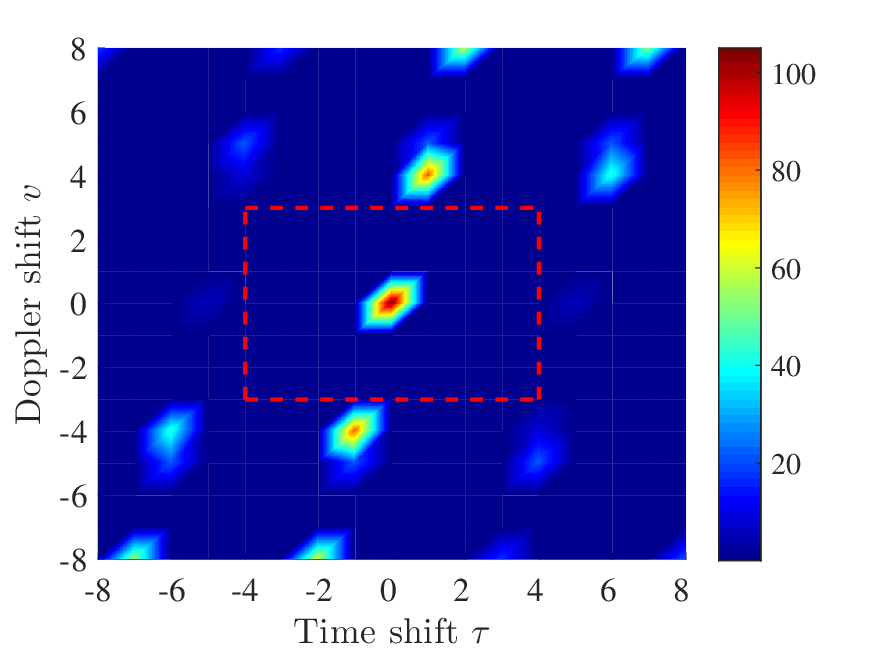}
 }
\subfigure[{The auto-ambiguity magnitudes of ${\bm{s}}_0$ over $[-4,4]\times [-3,3]$.}
]{
 \includegraphics[width=0.315\textwidth]{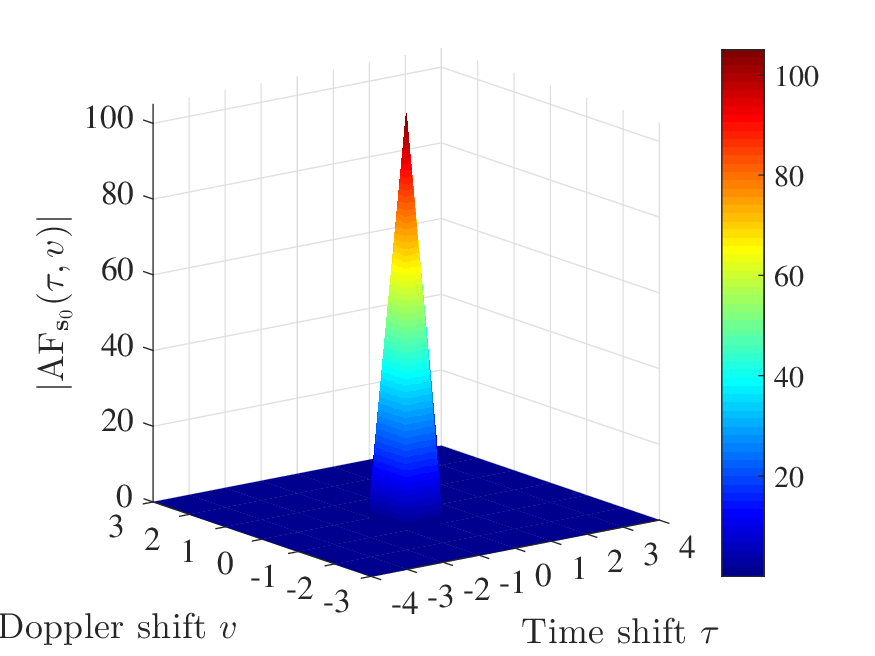}
 }
 \subfigure[{The cross-ambiguity magnitudes of ${\bm{s}}_0$ and ${\bm{s}_1}$ over $[-8,8]\times [-8,8]$.}
 ]{
 \includegraphics[width=0.315\textwidth]{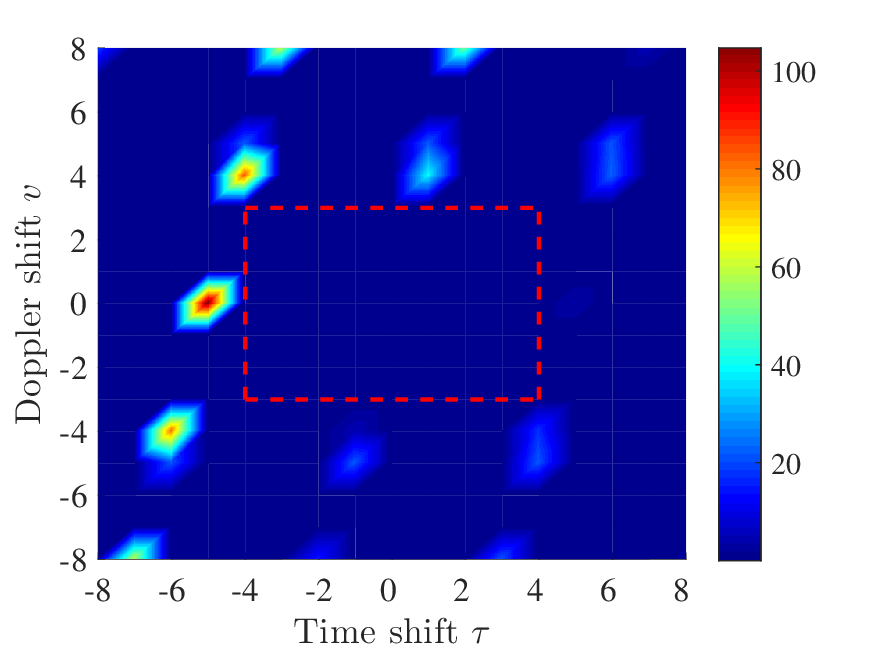}
 }
\caption{The ambiguity magnitudes of ${\bm{s}}_0$ and ${\bm{s}_1}$ in $\mathcal{S}$ from \emph{Example 2}.}
\end{figure*}

\emph{Remark 3:} The zero ambiguity zone ratio ${\rm{ZAZ_{ratio}}}$ of the constructed $\left( N( KN+P),N,\Pi \right)$-ZAZ sequence set $\mathcal{S}$ with $\Pi =(-N,N)\times (-K,K)$ is
\begin{align}
{\rm{ZAZ_{ratio}}}=1-\frac{P}{NK+P}.
\end{align}
Note that $\mathop {\lim}\limits_{NK\rightarrow \infty }{\rm{ZAZ_{ratio}}}\rightarrow 1$, indicating that the constructed ZAZ sequence set $\mathcal{S}$ is asymptotically optimal with respect to the theoretical bound in \emph{Lemma 2}.

{\emph{Example 2:}} Let $N=5$, $K=4$, and $P=1$.
Following \emph{Construction B}, a $(105,5,\Pi )$-ZAZ sequence set ${\mathcal{S}}=\left\{ {\bm{s}_n} \right\}_{n=0}^4$ with $\Pi =(-5,5)\times (-4,4)$ can be derived, where the $t$-th entry of ${\bm{s}}_n$ is given by
\begin{align*}
s_n(t)= \omega _{21}^{4t_1t_0}\cdot\omega _{5}^{nt_0},
\end{align*}
$0\leq t\leq 104$, $t_1=\lfloor t/5 \rfloor$, and $t_0={\langle t \rangle }_5$.
The zero ambiguity zone ratio of ${\mathcal{S}}$ is ${\rm{ZAZ_{ratio}}}=0.952381$.
The magnitudes of frequency-domain dual ${\bm{d}_n}$ corresponding to the sequence ${\bm{s}_n}$ are shown in Fig. 2, where $0\leq n\leq 4$.
Note that ${\bm{d}_n}$ has zero spectral power over the frequency-index set $\Omega=\left\{21\alpha+4\beta+\gamma: \alpha,\beta\in\mathbb{Z}_5,\,\gamma\in\mathbb{Z}^*_4\right\}\cup\left\{20 + 21 \alpha: \alpha \in \mathbb{Z}_5\right\}$.
It means that the sequence set $\mathcal{S}$ satisfies the spectral-null constraint $\Omega$.
The auto-ambiguity magnitudes of the sequence $\bm{{s}}_0$ over $[-8,8]\times [-8,8]$ and $[-4,4]\times[-3,3]$, and the cross-ambiguity magnitudes of the sequences ${\bm{s}_0}$ and ${\bm{s}_1}$ over $[-8,8]\times [-8,8]$ are shown in Fig 3. (a), Fig 3. (b), and Fig 3. (c) respectively.
It can be seen  that ${\bm{s}_0}$ has zero auto-ambiguity sidelobes over $[-4,4]\times[-3,3]$, ${\bm{s}_0}$ and ${\bm{s}_1}$ have zero cross-ambiguity magnitudes over $[-4,4]\times[-3,3]$.

{\emph{Remark 4:} In a cognitive radio/radar system, sequences are required to satisfy a spectral constraint such that zero or very low transmit power is allocated to certain forbidden carriers which are reserved for primary user(s) [32], [35], [36].
In [19], the energy gradient method and the Hu-Liu algorithm were combined to jointly optimize the local auto-ambiguity functions as well as the peak-to-average power ratio (PAPR) for a sequence under arbitrary spectral constraint.
In [14], transform domain approaches were proposed for generating sequences with low ambiguity magnitudes.
Unfortunately, [14] does not guarantee a constant modulus constellation for the entries of the generated sequences and this may result in a high PAPR.
In this section, a class of polyphase ZAZ sequence sets is derived in \emph{Theorem 3}, which have ideal PAPR and zero spectral power over certain spectral constraint $\Omega$.
These sequences are useful in cognitive communication and radar systems operating over certain  non-contiguous carriers}

\section{Proposed Construction of LAZ Sequence Sets}

In this section, we introduce a construction of asymptotically optimal periodic LAZ sequence sets based on a class of novel mapping functions.

\emph{Construction C:}

Let $p$ be an odd prime, $\pi:\mathbb{Z}_{p-1} \rightarrow \mathbb{Z}_p$ be a mapping function such that for any $a\in \mathbb{Z}^*_{p-1}$ and $b\in \mathbb{Z}_p$, $\pi({\langle x+a\rangle}_{p-1})\equiv \pi(x)+b\,\mathrm{mod}\, p$ has at most one solution for $x\in \mathbb{Z}_{p-1}$.
Construct a sequence set ${\mathcal{S}}=\left\{{\bm{s}_n} \right\}_{n=0}^{p-1}$ containing $p$ sequences of length $p( p-1 )$.
The $t$-th entry of ${\bm{s}_{n}}$ is given by
\begin{align}
s_n(t)=\omega _p^{t_1\pi(t_0)+n t_0},
\end{align}
where $0\leq t\leq p(p-1)-1$, $t=(p-1)t_1+t_0$, $t_1=\lfloor t/(p-1) \rfloor $, and $t_0={\langle t \rangle }_{p-1}$.

{\emph{Theorem 4:}} The sequence set $\mathcal{S}$ constructed above has the following properties:
\begin{enumerate}
\item It is a $p$-ary $\left(p(p-1),p,p,\Pi\right)$-LAZ sequence set with $\Pi =(-p+1,p-1) \times (-p,p)$;
\item Each sequence is an LAZ sequence with the maximum auto-ambiguity magnitude $p$ over the delay-Doppler zones $(-p+1,p-1) \times (-p(p-1),p(p-1))$ and $(-p(p-1),p(p-1) )\times (-p,p)$;
\item All the sequences in $\mathcal{S}$ are cyclically distinct.
\end{enumerate}

{\emph{Proof:}} 1) We first show that the sequence set $\mathcal{S}$ has low ambiguity
properties over a delay-Doppler zone around the origin.
Let $\bm{s}_{n}$ and $\bm{s}_{n'}$ be any two sequences in $\mathcal{S}$, where $0\leq n,n'\leq p-1$.
Calculate the periodic ambiguity function of $\bm{s}_{n}$ and $\bm{s}_{n'}$ as follows:
\begin{align}
&{\rm{AF}}_{{\bm{s}_n}, {\bm{s}_{n'}}} ( \tau ,v )\notag\\
=&\sum_{t=0}^{p(p-1)} s_n(t) \cdot s^*_{n'} ( \langle t+\tau \rangle_{p(p-1)}) \cdot \omega _{p(p-1)}^{vt} \notag\\
=&\sum_{t_1=0}^{p-1} \sum_{t_0=0}^{ p-2 } \omega _p^{t_1 \pi(t_0)} \cdot \omega _p^{nt_0} \cdot \omega _{p}^{-(t_1+\tau_1+ \delta_{t_0, \tau_0}) \pi({\langle t_0 + \tau_0\rangle }_{p-1})}\notag\\
&
\cdot \omega_p^{-n' ( t_0 + \tau_0 - (p-1) \delta_{t_0,\tau_0})} \cdot \omega _{p(p-1 )}^{v((p-1)t_1+t_0)} \notag\\
=& \sum_{t_0=0}^{p-2} \omega_p^{( n-n' )t_0}\cdot \omega_p^{-(\tau_1+\delta_{t_0,\tau_0} ) \pi( {\langle t_0 + \tau_0\rangle}_{p-1})} \cdot \omega _p^{-n' (\tau_0+\delta_{t_0,\tau_0})} \notag\\
&\cdot \omega_{p(p-1)}^{vt_0} \cdot \sum_{t_1=0}^{p-1} {\omega_p^{(\pi(t_0)- \pi({\langle t_0 + \tau_0\rangle }_{p-1})+v)t_1}},
\end{align}
where $t_1=\lfloor {t}/{(p-1)} \rfloor $, $t_0={\langle t \rangle }_{p-1}$, $\tau=(p-1) \tau_1 + \tau_0$, $\tau_1=\lfloor {\tau}/{(p-1)} \rfloor $, $\tau_0={\langle \tau \rangle}_{p-1}$, and $\delta_{t_0,\tau_0} = \lfloor( t_0+\tau_0)/(p-1) \rfloor $.

Consider the following four cases:

{\emph{Case 1:}} When $ \tau_0\neq 0$, there is at most one solution $t'_0$ with $0\le {t'_0}\le p-2$ such that $\pi({\langle t_0+\tau_0\rangle}_{p-1}) \equiv \pi(t_0)+v\,\mathrm{mod}\, p$.
If $\pi({\langle t_0+\tau_0\rangle }_{p-1}) \not\equiv \pi(t_0)+v\,\mathrm{mod}\, p$, $\sum_{t_1=0}^{p-1} \omega_p^{( v+\pi(t_0)-\pi({\langle t_0+\tau_0\rangle}_{p-1}))t_1}=0$ holds in (39), which follows that ${\rm{AF}}_{s_n,s_{n'}}(\tau,v )=0$.
Otherwise, there is a solution $t'_0$ with $0\le {t'_0}\le p-2$ such that $\pi({\langle t_0+\tau_0\rangle }_{p-1})\equiv \pi(t_0) +v\,\mathrm{mod}\, p$,
then
\begin{align}
 \left| {\rm{AF}}_{{\bm{s}_n},{\bm{s}_{n'}}}(\tau,v) \right|=&\left| p \cdot \omega_p^{(n-n') t'_0}\cdot\omega _p^{-(\tau_1+\delta_{t'_0,\tau_0}) \pi({\langle t'_0+\tau_0\rangle }_{p-1})} \right.\notag\\
&\left. \cdot \omega_p^{-n'(\tau_0-(p-1) \delta_{t'_0,\tau_0})} \cdot \omega _{p(p-1)}^{vt'_0} \right|\notag\\
=&p.
\end{align}

{\emph{Case 2:}} When $\tau_0 =0$ and ${\langle v \rangle }_p\neq 0$, we have
\begin{align}
 & {\rm{AF}}_{{\bm{s}_n},{\bm{s}_{n'}}}(\tau,v ) \notag\\
=& \sum_{t_0=0}^{p-2} \omega_p^{(n-n')t_0}
\cdot \omega_p^{-\tau_1 \pi({t_0})}\cdot \omega_{p(p-1)}^{vt_0} \cdot \sum_{t_1=0}^{p-1}{\omega_p^{vt_1}}\notag\\
=&0,
\end{align}
where $\sum_{t_1=0}^{p-1}{\omega_p^{vt_1}}=0$ for ${\langle v \rangle }_p\neq 0$.

{\emph{Case 3:}} When $n=n'$, $\tau=0$, ${\langle v \rangle }_p=0$, and $v\ne 0$, suppose $v=rp$, where $0< |r|<p-1$, then (39) reduces to
\begin{align}
{\rm{AF}}_{{\bm{s}_n}}(0,v)&=p\cdot \sum_{t_0=0}^{p-2}\omega_{p-1}^{rt_0}=0.
\end{align}

{\emph{Case 4:}} When $n\ne n'$, $\tau=0$, and $v=0$, (39) becomes
\begin{align}
{\rm{AF}}_{{\bm{s}_n},{\bm{s}_{n'}}}(0,0)&=p\cdot\sum_{t_0=0}^{p-2}\omega_p^{(n-n')t_0}=0,
\end{align}
where $0<|n-n'|\leq p-1$.

Combining {\emph{Case 1}} and {\emph{Case 2}}, we assert that the auto-ambiguity magnitude $\left|{\rm{AF}}_{\bm{s}_n}(\tau,v)\right|\leq p$ for $|\tau|< p(p-1)$, $|v|< p$, and $(\tau,v)\neq (0,0)$.
Combining {\emph{Case 1}}, {\emph{Case 2}}, and {\emph{Case 3}},
we observe that the auto-ambiguity magnitude $\left|{\rm{AF}}_{\bm{s}_n}(\tau,v)\right|\leq p$ for $|\tau|< p-1$, $|v|< p(p-1)$, and $(\tau,v)\neq (0,0)$.
Consequently, each sequence has the maximum auto-ambiguity magnitude $p$ over the delay-Doppler zones $(-p+1,p-1)\times (-p(p-1),p(p-1))$ and $(-p(p-1),p(p-1))\times (-p,p)$.
Combining {\emph{Case 1}}, {\emph{Case 2}}, and {\emph{Case 4}}, we have that the maximum cross-ambiguity function $|{\rm{AF}}_{{\bm{s}_n},{\bm{s}_{n'}}}(\tau,v)|< p$ for $|\tau|< p-1$, $|v|< p$, and $n\neq n'$.
Then, it is sufficient to show that the sequence set $\mathcal{S}$ is a $\left( p( p-1 ),p,p,\Pi\right)$-LAZ sequence set with the maximum periodic ambiguity magnitude $p$ over the delay-Doppler zone $\Pi=(-p+1,p-1)\times (-p,p)$.

2) Next, we show that all the sequences in $\mathcal{S}$ are cyclically distinct.
Assume on the contrary that $\bm{s}_n$ and $\bm{s}_{n'}$ with $0\leq n,n'\leq p-1$ in $\mathcal{S}$ are cyclically equivalent at the time shift $\tau$.
It implies that
\begin{align}
s_n(t)= s_{n'}(\langle{t + \tau \rangle }_{p(p-1)})\cdot \omega_p^c
\end{align}
holds for all $0\leq t \leq p(p-1)-1$, where $c \in\mathbb{Z}_p$.
It follows from (38) that for all $0\leq t_1 \leq p-1$ and $0\leq t_0 \leq p-2$, there is
\begin{align}
&(\pi(t_0) - \pi ({\langle t_0 + \tau_0 \rangle }_{p-1}) )t_1 + (n-n')t_0 \equiv n'(\tau_0+ \delta _{t_0,\tau_0}) \notag\\
& +(\tau_1 + \delta_{t_0,\tau_0})\pi (\langle {t_0 + \tau_0} \rangle_{p - 1})+ c\,\bmod \,p.
\end{align}
Note that for any $0\leq t_1\leq p-1$, (45) holds if and only if $\pi(t_0)- \pi ({\langle t_0 + \tau_0 \rangle }_{p - 1})=0$.
Thus we have $\tau_0 = 0$ and $\delta_{t_0,\tau_0} = 0$, and then it follows from (45) that
\begin{align}
(n - n')t_0 \equiv \tau_1\pi (t_0) + c\,\bmod \,p
\end{align}
holds for all $0\leq t_0 \leq p-2$.
Since $n\neq n'$, we have $\tau_1\neq 0$, and then
\begin{align}
\pi (t_0) \equiv\frac{n - n'}{\tau_1}t_0-\frac{c}{\tau_1}\,\bmod \,p.
\end{align}
According to (47), for any $a\in \mathbb{Z}^*_{p-1}$, there is $\pi (\langle t_0 + a\rangle_{p - 1}) =\pi (t_0) + \frac{n-n'}{\tau_1}a\,\bmod \,p$ for $0 \le t_0  \leq p-2$, which contradicts with the definition of $\pi$ in \emph{Construction C}.
Consequently, we deduce that all the sequences in $\mathcal{S}$ are cyclically distinct.

It is noted that \emph{Construction C} builds a connection between a class of mapping functions and the associated LAZ sequences.
The key of this construction is to find suitable mapping functions $\pi$ that satisfy the condition in \emph{Construction C}.
The following lemma presents such a class of mapping functions.

\emph{Lemma 4:} For an odd prime $p$, let $\pi(x)={\alpha}^{x}$ be a mapping function from $\mathbb{Z}_{p-1}$ to $\mathbb{Z}^*_p$, where $\alpha$ is a primitive element of $\mathbb{F}_p$.
For any $a\in \mathbb{Z}^*_{p-1}$ and $b\in \mathbb{Z}_p$, $\pi({\langle x+a\rangle}_{p-1})\equiv \pi(x)+b\,\mathrm{mod}\, p$ has at most one solution for $x\in \mathbb{Z}_{p-1}$.

\emph{Proof:} When $b=0$, the equation $\pi({\langle x+a \rangle}_{p-1})- \pi(x) = {\alpha}^x (\alpha^a-1) = 0$ has no solution for $x\in \mathbb{Z}_{p-1}$ as $a\in \mathbb{Z}^*_{ p-1 }$.
When $b\in \mathbb{Z}^*_p$, the equation $\pi({\langle x+a \rangle} _{p-1}) - \pi(x)={\alpha}^x ( \alpha^a-1 ) =b$ has exactly one solution for $x\in \mathbb{Z}_{p-1}$.
The proof of this lemma is then completed.

It might be possible and interesting to obtain more mapping functions $\pi:\mathbb{Z}_{p-1} \rightarrow \mathbb{Z}_{p}$ that satisfy the condition in \emph{Construction C} other than the one mentioned in \emph{Lemma 4}. The reader is kindly invited to search such mapping functions.

{\emph{Remark 5:}} For the constructed $\left( p(p-1), p, p, \Pi \right)$-LAZ sequence set $\mathcal{S}$ with $\Pi =( -p+1, p-1 )\times (-p, p)$, the tightness factor is
\begin{align}
 \rho_{\rm{LAZ}}=\left(1 + \frac{1} {p - 1} \right)\sqrt {1 - \frac{1} {p(p - 1)}}
\end{align}
Note that $\mathop {\lim}\limits_{p \to \infty}\rho_{\rm{LAZ}}\rightarrow 1$,
indicating that the constructed LAZ sequence set $\mathcal{S}$ asymptotically achieves the theoretical lower bound in \emph{Lemma 1}.
Similarly, one can check that each LAZ sequence in $\mathcal{S}$ asymptotically achieves the theoretical lower bound as $p$ increases.

To further visualize the parameters of the constructed LAZ sequence sets,
some explicit values of the parameters are listed in Table \uppercase\expandafter{\romannumeral3}.
Since the optimality factor $\rho_{\rm{LAZ}}$ is a meaningful figure for measuring the merit of LAZ sequence sets, we also list it in this table.
The numerical results show that the optimality factor $\rho_{\rm{LAZ}}$ of the constructed LAZ sequence sets asymptotically achieves 1 as $p$ increases, which means that the constructed LAZ sequence sets are indeed asymptotically optimal as predicted in \emph{Remark 5}.

\begin{table}[!t]
\caption{Parameters of the proposed $\left(L,N,\Pi,\theta_{\mathrm{max}}\right)$-LAZ sequence set}\centering
\begin{tabular}{|c|c|c|c|c|c|}
\hline
$p$&$L$ &$N$ &$\Pi$&$\theta _{\textrm{max}}$ &$\rho_{\rm{LAZ}}$\\
\hline
3 &6 &3 &$(2,2)\times(3,3)$ &3 &1.369306\\
5 &20 &5 &$(4,4)\times(5,5)$ &5 &1.218349\\
7 &42 &7 &$(6,6)\times(7,7)$ &7 &1.152694\\
11 &110 &11 &$(10,10)\times(11,11)$ &11 &1.094989\\
13 &156 &13 &$(12,12)\times(13,13)$ &13 &1.079856\\
17 &272 &17 &$(16,16)\times(17,17)$ &17 &1.060545\\
19 &342 &19 &$(18,18)\times(19,19)$ &19 &1.054011\\
23 &506 &23 &$(22,22)\times(23,23)$ &23 &1.044421\\
29 &812 &29 &$(28,28)\times(29,29)$ &29 &1.035076\\
31 &930 &31 &$(30,30)\times(31,31)$ &31 &1.032778\\
37 &1332 &37 &$(36,36)\times(37,37)$ &37 &1.027392\\
41 &1640 &41 &$(40,40)\times(41,41)$ &41 &1.024687\\
\hline
\end{tabular}
\end{table}

\begin{figure*}[!t]
 \centering
\subfigure[{The auto-ambiguity magnitudes of ${\bm{s}}_0$.
 }]{
 \includegraphics[width=0.32\textwidth]{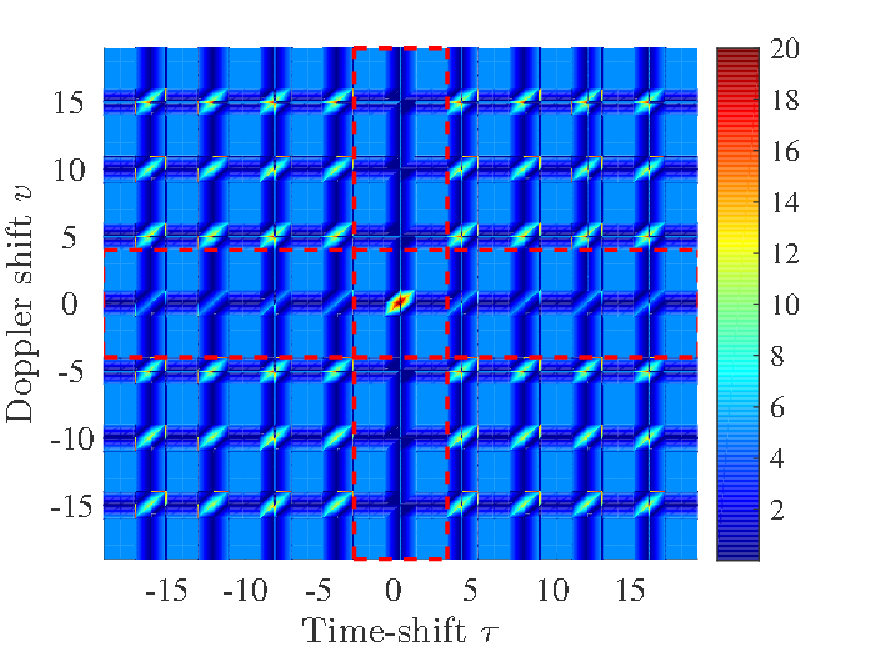}
 } \hspace{25mm}
\subfigure[{The auto-ambiguity magnitudes of ${\bm{s}}_0$ over $[-3,3]\times [-19,19]$.}
]{
 \includegraphics[width=0.32\textwidth]{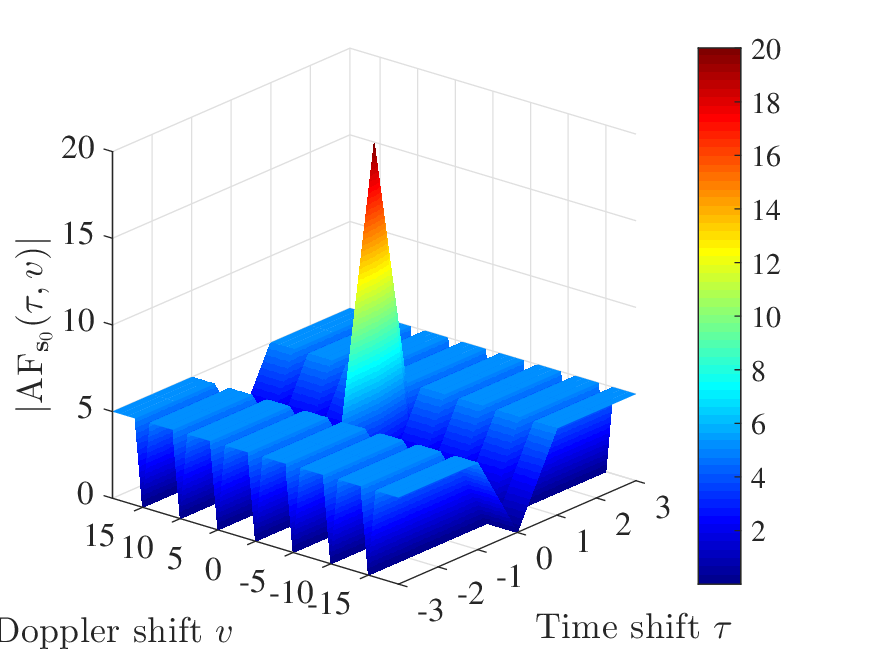}
 }
 \subfigure[{The auto-ambiguity magnitudes of ${\bm{s}}_0$ over $[-19,19]\times [-4,4]$.}
]{
 \includegraphics[width=0.32\textwidth]{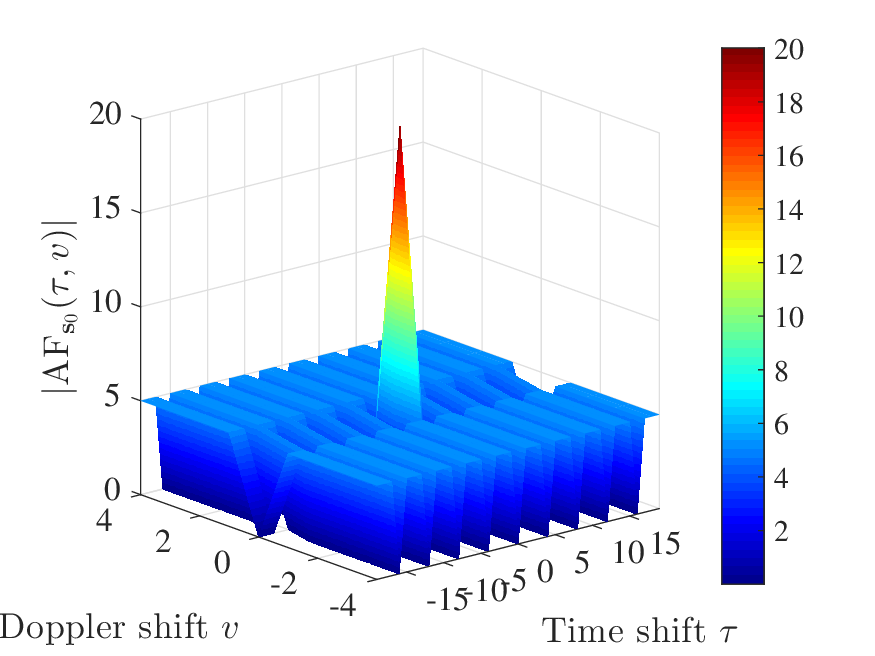}
 }\hspace{25mm}
 \subfigure[{The cross-ambiguity magnitudes of ${\bm{s}}_0$ and ${\bm{s}_1}$.}
 ]{
 \includegraphics[width=0.32\textwidth]{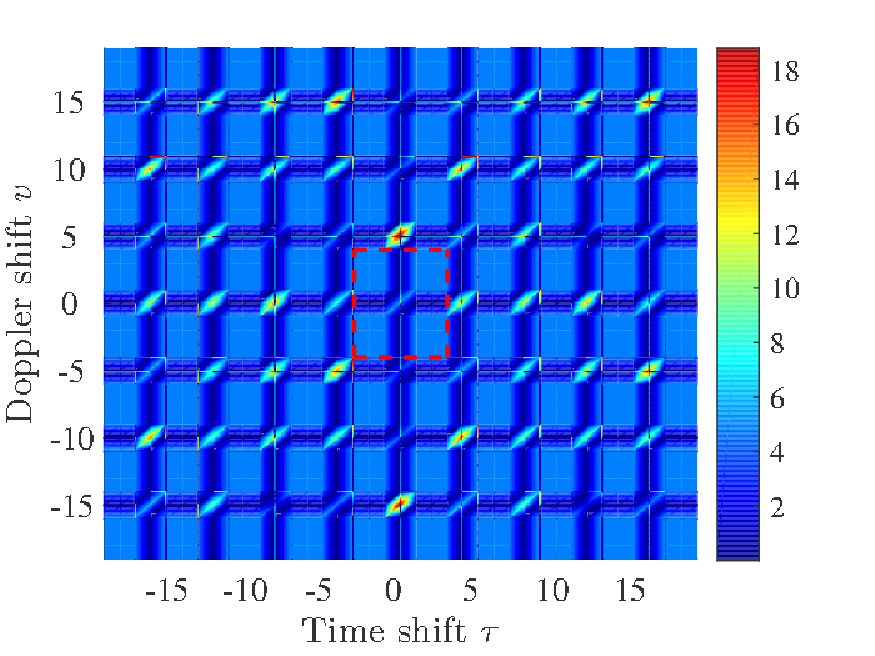}
 }
\caption{The ambiguity magnitudes of ${\bm{s}}_0$ and ${\bm{s}_1}$ in $\mathcal{S}$ from \emph{Example 3}.}
\end{figure*}

An example is given below to illustrate the proposed construction.

{\emph{Example 3:}} Let $p=5$ and $\pi(x)={\alpha}^{x}$, where $\alpha=3$ is a primitive element of $\mathbb{F}_5$, $x\in \mathbb{Z}_{4}$.
Following \emph{Construction C}, a sequence set $\mathcal{S}=\left\{ {\bm{s}_n}\right\}_{n=0}^4$ with each sequence of length 20 can be derived, where the $t$-th entry of ${\bm{s}_n}$ is given by
\begin{align*}
s_n(t)=\omega_5^{t_1 \cdot\pi(t_0)+nt_0},
\end{align*}
$0\leq t\leq 19$, $t_1=\lfloor {t}/4 \rfloor$, and $t_0={\langle t \rangle }_4$.
The sequences in $\mathcal{S}$ are listed as follows, where each element represents a power of $\omega_{5}$.
\begin{align*}
\bm{s}_0=
(&0,0,0,0,1,3,4,2,2,1,3,4,3,4,2,1,4,2,1,3);\\
\bm{s}_1=
(&0,1,2,3,1,4,1,0,2,2,0,2,3,0,4,4,4,3,3,1);\\
\bm{s}_2=
(&0,2,4,1,1,0,3,3,2,3,2,0,3,1,1,2,4,4,0,4);\\
\bm{s}_3=
(&0,3,1,4,1,1,0,1,2,4,4,3,3,2,3,0,4,0,2,2);\\
\bm{s}_4=
(&0,4,3,2,1,2,2,4,2,0,1,1,3,3,0,3,4,1,4,0).
\end{align*}
One can verify that $\mathcal{S}$ is a $(20,5,5,\Pi)$-LAZ sequence set with $\Pi=(-4,4)\times (-5,5)$ and the optimality factor $\rho_{\rm{LAZ}}=1.218349$.
The auto-ambiguity magnitudes of $\bm{s}_0$ over $[-19,19]\times [-19,19]$, $[-3,3]\times [-19,19]$, and $[-19,19]\times [-4,4]$, and the cross-ambiguity magnitudes of $\bm{s}_0$ and $\bm{s}_1$ over $[-19,19]\times [-19,19]$ are shown in Fig 4. (a), Fig 4. (b), Fig 4. (c), and Fig 4. (d) respectively.
It can be seen that $\bm{s}_0$ has the maximum auto-ambiguity sidelobe 5 over $[-3,3]\times [-19,19]$ and $[ -19,19]\times [-4,4]$, $\bm{s}_0$ and $\bm{s}_1$ have the maximum cross-ambiguity magnitude 5 over $[-3,3]\times [-4,4]$.

\section{Conclusions}

This paper is devoted to developing novel unimodular sequence sets with interesting ZAZ and LAZ properties.
We have first proposed two classes of polyphase ZAZ sequence sets in \emph{Construction A} and \emph{Construction B}, whereby the zero ambiguity sidelobes are obtained 1) by generalizing the PNF induced ZCZ construction in [29] and 2) by introducing successive nulls in the sequence frequency-domain, respectively.
Besides, a class of polyphase LAZ sequence sets has been presented in \emph{Construction C} with the aid of a novel class of mapping functions introduced in \textit{Lemma 4}.
These proposed sequence sets have been proven to be cyclically distinct and asymptotically optimal.

Due to low/zero ambiguity functions over a delay-Doppler zone around the origin, LAZ/ZAZ sequences have potential applications in future high-mobility communications systems, satellite networks, and radar sensing systems.
It is interesting to apply the proposed LAZ/ZAZ sequences in these systems to examine the relevant communication/sensing gains in various practical settings.
New optimal or asymptotically optimal LAZ/ZAZ sequences with more flexible parameters are also expected.

\section*{Acknowledgment}

The authors would like to thank anonymous reviewers and the Associate Editor Dr. Gohar Kyureghyan for their valuable comments and suggestions that much improved the quality of this paper.

\end{document}